# A computational theory for the production of limb movements

**Short title:** Optimal motor control

Emmanuel Guigon (https://orcid.org/0000-0002-4506-701X)

Sorbonne Université, CNRS, Institut des Systèmes Intelligents et de Robotique, ISIR, F-75005 Paris, France

**Corresponding author:**

Emmanuel Guigon

Sorbonne Université, CNRS, Institut des Systèmes Intelligents et de Robotique, ISIR

Pyramide Tour 55 - Boîte Courier 173 - 4 Place Jussieu,75252 Paris Cedex 05, France

Tel: 33 1 44276382 / Fax: 33 1 44275145 / Email: emmanuel.guigon@sorbonne-universite.fr

**Author Note**: The ideas in this article are inspired from a preceding article (Guigon et al., 2019), but none of the material presented here has been presented in a symposium or a meeting, nor posted on preprint server. The author declares no competing financial interests.

# Abstract

Motor control is a fundamental process that underlies all voluntary behavioral responses. Several different theories based on different principles (task dynamics, equilibrium-point theory, passive-motion paradigm, active inference, optimal control) account for specific aspects of how actions are produced, but fail to provide a unified view on this problem. Here we propose a concise theory of motor control based on three principles: optimal feedback control, control with a receding time horizon, and task representation by a series of via-points updated at fixed frequency. By construction, the theory provides a suitable solution to the degrees-of-freedom problem, i.e. trajectory formation in the presence of redundancies and noise. We show through computer simulations that the theory also explains the production of discrete, continuous, rhythmic and temporally-constrained movements, and their parametric and statistical properties (scaling laws, power laws, speed/accuracy tradeoffs). The theory has no free parameters and only limited variations in its implementation details and in the nature of noise are necessary to guarantee its explanatory power.

# Introduction

Action is the only mean by which the nervous system can communicate evolutions of its internal states to the external world. In any case and irrespective of any theoretical construct, to produce a faint smile, a friendly handshake or a full running pattern, a voluntary (as opposed to a reflex) process should be triggered to create an appropriate temporal pattern of coordination directed to specific muscular groups (Lashley, 1951; Bernstein, 1967). Each and everyone could agree on this statement, yet there is no consensus on the nature of this process (not to say on its anatomical and physiological bases; Arber & Costa, 2018). Although formal debates on this issue has long since disappeared from the literature (Feldman & Levin, 1995; Kelso, 1995; Turvey, 1977; but for some recent revivals, see Friston, 2011; Huys, Perdikis, & Jirsa, 2014; Mohan, Bhat, & Morasso, 2019), one is still confronted with the same recurring and embarrassing questions when addressing motor control: Are there internal representations in the brain? Does one part of the brain act as a controller which enslaves other parts of the brain and the body? Does the brain manipulate position or force variables? In this framework, a central issue is as follows: how not to keep contemplating and discussing these difficulties, and make progress in the field of motor control modeling that could be beneficial for understanding movement disorders, improving rehabilitation devices or inspiring human-like robotics?

The goal of this article is to describe a computational theory[1] for the production of limb movement that resolves or circumvents part of these difficulties. The starting point is a recent study of velocity fluctuations during slow movements (Guigon, Chafik, Jarrassé, & Roby-Brami, 2019). This study shows that a slow movement (with a mean speed below 10 cm/s) is well described by a time series of constant-duration (~0.13 s), elementary, discrete displacements whose amplitude is proportional to mean movement speed[2]. The central observation is that the temporal structure of movement kinematics is invariant with changes in movement speed, consistent with the landmark study of Vallbo and Wessberg (1993). The study proposes a model to explain how slow movements are produced: a regular staircase goal position signal at ~8 Hz is pursued by an optimal feedback controller with a temporal horizon of 0.28 s[3]. To visualize the process, imagine that you try to reach a virtual target that jumps to a different place every 1/8 s and that your strategy is to assume that at each time the target is stationary and should be reached smoothly in 0.28 s. The resulting movement will be a series of aborted smooth segments of 1/8 s whose velocity is proportional to the size of jumps. The model suggests a general theoretical account of motor control in terms of goals intermittently updated at this frequency and optimally pursued at this horizon. According to the model, any fluctuation observed in a kinematic signal (diversely called submovement, segment, unit, pulse, ...) should be considered a consequence of the pursuit of a temporary goal[4]. Here, a theory is derived on this basis in terms of three computational principles, and is shown to explain a broad range of motor phenomena: trajectory formation in discrete (point-to-point), continuous (drawing, handwriting), rhythmic and temporally-constrained tasks, ubiquity of isochronous behaviors, scaling laws, power laws and speed-accuracy tradeoffs. The article is organized in four parts. First, we outline the context of this study. Next, we present our computational theory of motor control, and a set of simulations that illustrate its main characteristics. Finally, we provide a thorough discussion of the results and the theory.

---

[1] The term "computational" is used to designate a class of models based on the theory of control (including optimal control) and internal models (Todorov & Jordan, 2002; Wolpert & Ghahramani, 2000). Detailed reading of Todorov and Jordan (2002) is an important prerequisite to understand the present article.

---

[2] An illustrative representation is a regular temporal staircase signal with speed-dependent stair height.

[3] The notions of elementary displacement and temporal horizon are not ambiguous. They emerge from processing and model-based description of experimental data (Guigon et al., 2019). Yet the reported values (0.13 s for the duration of elementary displacements, 0.28 s for the temporal horizon) are not universal as they were obtained from data analysis and depend on the choice of a cutoff frequency for low-pass filtering.

[4] Important notions related to submovement and intermittency are introduced and comprehensively discussed in Guigon et al. (2019).

# Context

## The computational approach

The computational approach to motor control[5] is prototypically represented by the stochastic optimal feedback control (SOFC) theory of Todorov and Jordan (2002) which has reached a wide audience due to its conciseness, scope and explanatory power (Diedrichsen, Shadmehr, & Ivry, 2010; Scott, 2004). This theory provides a principled solution to the problem of motor coordination (Bernstein's degrees-of-freedom problem), i.e. how redundant degrees of freedom at each level of the motor hierarchy (from neural space to task space) are coordinated to produce goal-directed actions, and accounts for many characteristics of motor acts (trajectories, structured variability, synergy formation, ...) as by-products of its principles.

SOFC came on top of fifty years of research that elaborated on the application of control, optimal control and optimal estimation theory to the description of human motor behaviors (Baron & Kleinman, 1969; Flash & Hogan, 1985; Hatze, 1976; Harris & Wolpert, 1998; Hoff, 1994; Hogan, 1984; Nelson, 1983; Uno, Kawato, & Suzuki, 1989; Wolpert, Ghahramani, & Jordan, 1995). It contained not only a synthesis of previous proposals but also several new ideas on the nature of motor control. First, it abolished the long-held but embarrassing dichotomy between feedforward and feedback control, and replaced it by a more representative control/estimation architecture[6]. Second, it proposed to resolve

---

[5] See Discussion for a presentation of debates on motor control.
[6] In traditional control architectures, a feedforward controller generates an open-loop control signal (e.g. to follow a desired trajectory) and a feedback controller (e.g. a reflex) corrects

redundancy on a moment-by-moment basis and simultaneously for all the available degrees of freedom (from neural space to task space). Third it gave a central, instrumental role to signal-dependent motor noise (multiplicative noise on motor commands; Harris & Wolpert, 1998; Jones, Hamilton, & Wolpert, 2002; Todorov, 2002) in the emergence of a "minimum intervention principle" according to which "deviations from the average trajectory are corrected only when they interfere with task performance" (Todorov & Jordan, 2002). A likely consequence of this principle is the emergence of structured patterns of variability in which a large variability in the contribution of individual degrees of freedom from trial to trial can accompany a low goal-related variability (Bernstein's idea of "repetition without repetition"; concept of uncontrolled manifold; Scholz & Schöner, 1999).

## Limitations

Despite its successes, SOFC is plagued by several limitations, four of them being quite revealing. First, it comes with a heavy computational burden all the more so that a specific controller must be built for each task at hand. For instance, a via-point experiment (e.g. Experiment 1 in Todorov & Jordan, 2002) would require a dedicated controller for every possible configuration of the via-points. The burden results from the formulation of the model in terms of a Linear Quadratic Gaussian (LQG) controller in which the task representation is embedded into the cost function. This approach blurs the distinction between skilled and unskilled actions, and leaves no room for multiple, eventually

---

deviations due to perturbations (Schaal, Ijspeert, & Billard, 2003). Based on arguments drawn from observations on task variability and goal-directed motor corrections, Todorov and Jordan (2002) rejected these architectures and proposed a control scheme in which a feedforward command is elaborated at each time from feedback information (actual and predicted sensory inputs). The combination of feedforward and feedback control offers no room for open-loop control since control is recalculated at each time based on feedback information.

suboptimal strategies which can exist to solve a motor task (Ganesh, Haruno, Kawato, & Burdet, 2010; Kodl, Ganesh, & Burdet, 2011).

Second, as the task representation is embedded into the cost function, LQG runs several objectives in parallel (e.g. find a solution of lowest energy expenditure possible and the closest possible to a goal), and makes compromises between the objectives. Parameters are needed to weight the different objectives which are not easy to identify experimentally (how one would trade energetic expenditure for accuracy, or positional accuracy for velocity accuracy?), if they ever exist.

Third, as the time to achieve a goal is chosen in advance and decreases gradually as the goal draws nearer, continuous motor behaviors (e.g. tracking, drawing, writing, scribbling, …) and flexible adjustments following perturbations (Liu & Todorov, 2007; Shadmehr & Mussa-Ivaldi, 1994) are not properly explained in this framework. Ad hoc online updating of time is feasible (Liu & Todorov, 2007), but lacks an underlying principle. In fact, time representation is a deep and ubiquitous issue. Considering again a via-point experiment, not only the overall task duration must be chosen, but also the time to reach each via-point. Experimental studies have shown that the temporal organization of reaching through via-points obeys to an isochrony principle, i.e. the transit time between successive points is almost constant (Flash & Hogan, 1985; Kodl, Ganesh, & Burdet, 2011). Isochrony is a venerable concept corresponding to a compensatory regulation of movement speed with amplitude to maintain movement duration approximately constant (Binet & Courtier, 1893; Bryan, 1892; Denier van der Gon & Thuring, 1965; Glencross, 1975; Lacquaniti, Terzuolo, & Viviani, 1983; Viviani & McCollum, 1983). The origin of isochrony is unknown[7] and has been rarely addressed in computational studies (Flash & Hogan, 1985; Flash, Meirovitch, & Barliya, 2013; Saito, Tsubone, & Wada, 2006). Incomplete compensation can be explained in optimal control models by a cost of time (Harris & Wolpert, 2006; Hoff, 1994; Shadmehr, Orban de Xivry, Xu-Wilson, & Shih, 2010). Yet, in these models, complete compensation and strict temporal invariance (isochrony) are irrational and would require an infinite contribution of the cost of time.

---

[7] Note that isochrony is found in models using classical feedback control and in related models (e.g. Bullock & Grossberg, 1988).

Fourth, SOFC (and in fact all optimal control models) produces smooth movements of any duration whereas the smoothness of experimentally recorded movements decreases with increasing movement duration (Salmond, Davidson, & Charles, 2017; Shmuelof, Krakauer, & Mazzoni, 2012; review in Guigon et al., 2019). The discrepancy is due to the intrinsically time-invariant nature of optimal trajectories generated by the models. In the same vein, optimal control provides no account for movement intermittency (Doeringer & Hogan, 1998; Guigon et al., 2019; Vallbo & Wessberg, 1993), and the constant rate of peak velocity, peak acceleration and peak jerk in motor behaviors (Guigon et al., 2019; Shmuelof et al., 2012; Vallbo & Wessberg, 1993).

## Three computational principles

Three principles are proposed to exploit the power of the computational approach (Todorov & Jordan, 2002), and overcome the above-mentioned limitations. They are derived from experimental and theoretical observations on slow movements which suggest that a motor task is represented by a time series of goals updated at a fixed frequency and optimally pursued at a fixed horizon (Guigon et al., 2019). The principles are described in the framework of control theory. Motor control is considered as a *control problem* in which the behavior of a *controlled object* is governed by a *controller* through a *control policy* and a set of *goals* to achieve. The 3 principles are the following:

# The control policy is a "universal" optimal feedback control policy

An optimal feedback control (OFC) policy is a function that takes as input the best estimated state of the controlled object (as given by an optimal state estimator), a goal state and a time to reach the goal, and provides the best control (relative to a cost function) that drives the controlled object to the goal in the given time (Bryson & Ho, 1975). In the most general (nonlinear, continuous-time) setting, an OFC policy $\boldsymbol{u}(t)$ to reach a goal $\boldsymbol{x}^G(t)$ (which is a function of time since it can change at any time) can be formally written at each time $t$

$$\boldsymbol{u}(t) = \boldsymbol{U}(t),$$

where $\boldsymbol{U}$ is the function defined in $[t, T(t)]$ by

$$\boldsymbol{U}(t') = argmin_{\boldsymbol{u}'} \int_{t}^{T(t)} L(\boldsymbol{x}(\theta), \boldsymbol{u}'(\theta)) \, d\theta \quad \text{(Equation 1)}$$

for a dynamics of the controlled object given by

$$\dot{\boldsymbol{x}}(t) = \boldsymbol{f}(\boldsymbol{x}(t), \boldsymbol{u}(t)) + \boldsymbol{n}_{dyn}(t), \quad \text{(Equation 2)}$$

where $\boldsymbol{x}$ is the state of the controlled object[8] (*italic* is used for scalars, ***bold italic*** for vectors, and **bold** for matrices, dot is for time derivative), $T$ the time to reach the goal (which is considered a function of time; see below for the choice of $T$), $\boldsymbol{f}$ the dynamics of the object, $\boldsymbol{n}_{dyn}$ a noise on the dynamics, and $L$ a cost function. Boundary conditions are given by

$$\boldsymbol{x}(t) = \widehat{\boldsymbol{x}}(t),$$

and

$$\boldsymbol{g}(\boldsymbol{x}(T(t)), \boldsymbol{x}^G(t)) = 0, \quad \text{(Equation 3)}$$

---

[8] The state of the object corresponds to the smallest possible subset of variables that are necessary to describe the behavior of the object (e.g. position, velocity, ...).

where $g$ is a function that specifies the final boundary conditions[9], and the best estimated state $\hat{x}$ is defined by

$$\dot{\hat{x}}(t) = f(\hat{x}(t), u(t)) + \mathbf{K}(t)(y(t) - \mathbf{O}\hat{x}(t)), \quad \text{(Equation 4)}$$

where $\mathbf{K}$ is the Kalman gain, $\mathbf{O}$ the observation matrix, and

$$y(t) = \mathbf{O}x(t) + n_{obs}(t), \quad \text{(Equation 5)}$$

where $n_{obs}$ is an observation noise term (see Stengel, 1984 for a mathematical description of optimal estimation and Kalman filtering). Equations 1 to 5 define an architecture that belongs to the class of control/estimation architectures (Figure 1A; Todorov, 2004). The fact that $x^G(t)$ is a function of time does not mean that it prescribes a "desired trajectory". The actual trajectory corresponding to $x^G(t)$ is not $x^G(t)$, but a concatenation of pieces of trajectory produced by the control policy to reach $x^G(t)$ at each time $t$.

In this formulation, the control policy (Equation 1) is deterministic and is the same irrespective of the task at hand, i.e. it is aware of the task only through constraints that set the goals (boundary conditions $x^G(t)$; Equation 3) to achieve for this task. Such a control policy is called "universal" to indicate that it is a general purpose process independent of any specific task, without any other connotations. In order to justify this choice, we must prove that this formulation has the same explanatory power as the stochastic formulation used by Todorov and Jordan (SOFC), and alleviates some limitations of SOFC.

*Figure 1*. Building the model. **A**. Control/estimation architecture corresponding to Equations 1-5. **B**. Normalized boxcar function. **C**. Example of a step function. Here, the update times are regularly spaced with step duration $T_{step}$.

---

[9] A usual function $g$ is $g(x(T(t)), x^G(t)) = x(T(t)) - x^G(t)$, but other functions can be used as well. Equation 3 indicates that the goal $x^G(t)$ is a goal state, which means that a goal can be possibly specified for every state variable (e.g. position, but also velocity, ...).

SOFC elaborates a stochastic control policy[10] which is optimal with respect to the statistics of the noise on the dynamics (Equation 2) and observation (Equation 5), based on

$$\boldsymbol{U}(t') = argmin_{\boldsymbol{u}'} \langle \int_{t}^{T(t)} L(\boldsymbol{x}(\theta), \boldsymbol{u}'(\theta), \boldsymbol{P}(\theta)) \, d\theta \rangle$$

where **P** is a set of parameters that specifies the task at hand, and ⟨ ⟩ the mathematical expectation operator over noise. In this framework, the task is represented by an objective (quantity to minimize) rather than by a constraint (condition to fulfill; Equation 3) (Nelson, 1983). Interestingly, the very mechanism which is responsible for the explanatory power of SOFC (Todorov & Jordan, 2002) is not univoquely related to a formulation as an LQG controller. The only requirements for the occurrence of the minimum intervention principle (see explanation above) are the presence of signal-dependent motor noise and a properly functioning optimal state estimator, irrespective of the stochastic or deterministic nature of the controller, and the formulation in terms of objectives or constraints (Guigon, Baraduc, & Desmurget, 2008b). More specifically, the fundamental observation that motor variability is organized along redundant task dimensions (Scholz & Schöner, 1999; Figure 1 in Todorov & Jordan, 2002) does not necessitate a full-blown formulation in terms of stochastic optimality. The mathematical formulation of SOFC requires a strikingly large number of terms (Equations 2 and 3 in Todorov & Jordan 2002, Supplementary Notes), but most of these terms could be removed without qualitative consequences on the results reported by Todorov and Jordan (Guigon et al., 2008b). Simulations have shown that a deterministic controller that handles control-dependent objectives and state-dependent, task-related constraints separately and a LQG controller account equally well for structured patterns of motor variability (Guigon et al., 2008b). The main difference between the two approaches is the fact that, in the former case, the controller is not stochastic, i.e. it is unaware of the statistics of noise. In fact, it remains to be proven that stochastic optimality is a necessary concept in motor control.

---

[10] The distinction between stochastic and deterministic processes concerns only the control policy. In all OFC formulations, optimal state estimation is present and stochastic by construction.

The proposed formulation (Equations 1 to 5) alleviates two limitations of SOFC. First, the computational burden of building a specific control policy for each task at hand is replaced by the burden of building a unique universal policy which is not more complex than any task specific policy[11]. Second, the parameters which are necessary to weight multiple simultaneous objectives in SOFC are absent when constraints are used rather than objectives, which eliminates extra parameters and extra rules to set them.

This discussion might appear uselessly technical. However, as long as principles are concerned, it is important to address the efficiency and conciseness of theoretical constructs. In practice, the two formulations discussed above have the same explanatory power.

## The time to reach a goal is constant irrespective of the time already spent for this goal

Except in highly specific laboratory conditions, neither the world nor the body should be considered as stationary, e.g. due to noise and uncertainties, there is no such thing as a fixed target or fixed posture. Accordingly there is no such thing as the onset, the middle or the end of an action, but only an ongoing state that continuously evolves toward ever changing goals. Temporal flexibility is necessary to account for these facts and is afforded by the function $T(t)$ (Equation 1). The typical choice $T(t) = T_0$, where $T_0$ is a constant (*finite fixed horizon*), offers no flexibility as the remaining time to achieve a goal at each time $t$ is $T_0 - t$ which decreases gradually as time passes and the goal draws nearer. In this framework, different $T_0$ are used to obtain movements of different durations which leads to the possibility of smooth movements of any duration in contradiction with the fact that movement smoothness decreases with increasing movement duration, and only the fastest movements are likely to be smooth. An alternative choice is $T(t) = \infty$ (*infinite horizon* formulation; Qian, Jiang, Jiang, & Mazzoni, 2013; Rigoux & Guigon, 2012). In this case, movement duration is an emergent characteristic of optimal control, which provides a principled solution to the problem of flexibility. Yet the problem of smoothness remains open. A third choice, retained here as a principle, is $T(t) = t + T_H$, where $T_H$ is a constant (*finite receding horizon*), which provides a constant time $T_H$ to achieve a goal at each time (Berio, Calinon, & Fol Leymarie, 2017; Bye & Neilson, 2008, 2010; Guigon et al., 2019). We use the self-explanatory term *receding horizon* for $T_H$[12]. In this case, the control policy becomes stationary (independent of time), produces isochronous behaviors, and is intrinsically flexible as any novel goal is automatically pursued at horizon $T_H$. Allowing $T_H$ to vary would lead to the abovementioned problem of smoothness. The receding horizon is thus considered fixed, not only within a movement, but in fact across all movements and tasks. The value of $T_H$ (0.28 s) was identified in a study of slow movements (Guigon et al., 2019). The most important point is not the value of $T_H$ by itself, but the fact that it must be considered a constant rather than an open parameter. How movements of different durations are obtained in the framework of the receding horizon is the object of the third principle (see below).

---

[11] The central problem of how an optimal control policy is built and stored by the nervous system remains unsolved.

## A task is defined by a sequence of goals played at a fixed and unique frequency

A strong proposal of Todorov and Jordan is that motor behavior is based on the pursuit of goals rather than tracking of prescribed trajectories. Accordingly, a task can be defined by a discrete sequence of successive, non overlapping goal states (Equation 3), e.g. a sequence of via-points (Flash & Hogan, 1985; Kodl et al., 2011; Experiment 1 in Todorov & Jordan, 2002; Wada & Kawato, 1995). In this framework, the most general definition of a task is a time series $\boldsymbol{x}^G(t)$ of $N$ goal states $\boldsymbol{x}_k^G$, updated at times $t_k$, given by

$$\boldsymbol{x}^G(t) = \sum_{k=1}^{N-1} \boldsymbol{x}_k^G \, \text{boxcar}(t, t_k, t_{k+1}) + \boldsymbol{x}_N^G \text{boxcar}(t, t_N, \infty) \quad \text{(Equation 6)}$$

for $t \in [t_1, \infty[$, where $\text{boxcar}(t, a, b)$ is the function which is 1 for $t \in [a, b]$ and 0 elsewhere (Figure 1B; https://en.wikipedia.org/wiki/Boxcar_function). The boxcar function allows a single goal to be selected at each time. We only consider time series with a constant time interval called $T_{step}$, i.e. $\forall k, t_{k+1} - t_k = T_{step}$. The resulting time series $\boldsymbol{x}^G(t)$ is a step function (https://en.wikipedia.org/wiki/Step_function), where each step corresponds to a goal that is pursued for a certain time (example in Figure 1C). The goal states are called *via-points* except the last one which is called *target*. If any possible task is to be described by Equation 6, a principle should be provided for the choice of goal states and update times (value of $T_{step}$) for a given task. The proposed principle is based on observations on slow movements (Guigon et al., 2019), and is summarized in five rules:

- <u>rule 1</u>: if the task involves state-space constraints (e.g. spatial constraints), the goal states are chosen to match the constraints. For instance, a tracking task would be represented by a series of goals extracted from the trajectory to follow;

- <u>rule 2</u>: if there are no state-space constraints (e.g. scribbling), the goal states are chosen at will;

- <u>rule 3</u>: if the task involves temporal constraints (e.g. follow the beat of metronome), the update times are chosen to match the constraints (e.g. Figure 1C with $T_{step}$ = period of the metronome);

- <u>rule 4</u>: if there are no temporal constraints (e.g. scribbling), the goal states are updated with a fixed period $T_G = 0.13 \, s$, i.e. $t_{k+1} - t_k = T_G$ (e.g. Figure 1C with $T_{step} = T_G$).

- rule 5: the presence of state-space and temporal constraints does not prevent from considering additional goal states updated with period $T_G$.

---

[12] The concept of receding horizon belongs to the framework of model predictive control (Garcia, Prett, & Morari, 1989). No technical information on this framework is needed to understand the model.

The central element of the proposed principle is in rule 4 which gives a unique recipe to specify the time course of a motor act without a direct specification of its duration. The actual duration is an emergent consequence of the interaction between $T_G$ and $T_H$, and an increasing function of $N$ (number of goal states). The principle states that $T_G$ is the same across all movements and tasks. The value of $T_G$ (0.13 s) was identified in Guigon et al. (2019), to account for velocity fluctuations at ~8 Hz during slow movements. As for $T_H$, the most important point is not the value of $T_G$ by itself, but the fact that it must be considered a constant rather than an open parameter (see Table 1 for a summary on time notations).

Note that when the goals are updated with the period $T_G$ and pursued at horizon $T_H$ (which is longer than $T_G$), a goal may not be reached before the occurence of the next one.

In the LQG formulation used by Todorov and Jordan, the sequence of goals is integrated in the cost function and the resulting trajectory corresponds to the best way to run through the sequence. In the present approach, the sequence of goals is decided upstream of actual control which allows for multiple, possibly suboptimal task representations.

# Simulations

We present a set of simulations that obey to the three proposed principles. They attempt to reproduce observed characteristics of motor behavior during different tasks. The architecture of the model (Figure 1A) is the same for all the simulations, and the simulations differ only by the sequence of goals used to describe the tasks (Figure 1C). The detailed functioning of the model is the following. A simulation duration $\Delta$ and a timestep of simulation $\delta$ are chosen. At each time $t$ in $[0, \Delta]$, a goal $x^G(t)$ and the estimated state $\hat{x}(t)$ are available and a control policy $u(t)$ is calculated over the interval $[t, t + T_H]$ to reach the current goal ($x^G(t)$) at time $t + T_H$ starting from the estimated state at time $t$ (Equations 1 and 3). Then the new state $x(t + \delta)$ and new estimated state $\hat{x}(t + \delta)$ at time $t + \delta$ due to the control, observation and noise at time $t$ are calculated from current state $x(t)$ and current estimated state $\hat{x}(t)$ (Equations 2, 4 and 5). The process is repeated at time $t + \delta$ and stopped when current time is $\Delta$. Note that the simulation duration $\Delta$ defines the period of time during which the behavior of the model is simulated and has no influence on this behavior. $\Delta$ is chosen to cover the duration of the sequence of goals (see Table 1 for a summary on time notations).

The principles do not specify a unique model. Some freedom remains on how sequences of goals are built (in allocentric or egocentric coordinates) and on the nature of constraints at each goal (e.g. position, velocity, force, ...). The possible variations of the model are systematically tested.

There are no free parameters in the deterministic simulations and only noise parameters in the stochastic simulations. The noise parameters are given with no justification (the noise parameters not specified are 0), but a specific section (the last one) is dedicated to the study of noise and variability.

## Methods

*One-dimensional inertial point*

The three computational principles were applied to an inertial point actuated by a linear muscle in the presence of noise, with a quadratic cost function, and a Kalman filter for optimal state estimation. The rationale for this choice (in particular the absence of redundancy) is the following. As the model inherits properties from the model of Todorov and Jordan (2002), we deem it not necessary to address issues already considered in their work (kinematic redundancy, muscular redundancy, formation of uncontrolled manifolds and synergies).

The dynamics of the point was given by Equation 2, with the state vector $\boldsymbol{x} = [p\ v\ a\ e]$ ($p$ position, $v$ velocity, $a$ muscular activation, $e$ muscular excitation), the control vector $\boldsymbol{u} = [u]$, and the vectorial function $\boldsymbol{f}$ defined by

$$\begin{cases} \dot{p} = v \\ m\dot{v} = a \\ \tau\dot{a} = -a + e \\ \tau\dot{e} = -e + u \end{cases}$$

where $m$ is the mass of the point and $\tau$ the muscle time constant. This formulation corresponds to a linear, second-order (Newtonian) dynamics coupled to a linear, second-order (low-pass filtering) force generator, and widely used in motor control models (Harris & Wolpert, 1998; Todorov & Jordan, 2002; van der Helm & Rozendaal, 2000).

The cost function was defined by

$$L(\boldsymbol{x}, \boldsymbol{u}) = u^2.$$

State estimation obeyed to Equation 4 and observation to Equation 5. The Kalman gain was calculated following Guigon et al. (2008b) for sources of noise described below.

*Task representation: series of goals and boundary conditions*

A task was represented by a time series of goal states (Equation 6). There are two ways to define a sequence. The next goal can be defined relative to the current goal or relative to the current (estimated) state of the system. It corresponds broadly to the distinction between allocentric and egocentric coding of goals. Thus we considered an *absolute goal setting* policy when

$$\boldsymbol{x}_k^G = \boldsymbol{x}_{k-1}^G + \boldsymbol{\alpha}_k \quad \text{(Equation 7)}$$

and a *relative goal setting* policy when

$$\boldsymbol{x}_k^G = \widehat{\boldsymbol{x}}(t_k) + \boldsymbol{\alpha}_k, \quad \text{(Equation 8)}$$

where $\boldsymbol{\alpha}_k$ is an arbitrary sequence.

In the most general setting, final boundary conditions were defined by Equation 3. The choice of function $g$ is a fundamental issue. For instance, different boundary conditions should probably be used for discrete and continuous movements, e.g. it may not be necessary or effective to impose a zero-velocity constraint at a via-point in a continuous movement. Yet there is no self-evident principle for the choice of $g$ (e.g. what is an appropriate velocity constraint at a via-point in a continuous movement?), and it is necessary to consider and test different possibilities. There are two aspects in the construction of boundary conditions. First, the states to be constrained are chosen, the remaining unconstrained states being automatically determined by the optimal control process. For a 4 dimensional problem (position, velocity, activation, excitation), there are 15 different configurations, but only three were considered:

- the *full-state* constraint, i.e.

$$\begin{cases} p(t+T_H) - p^G(t) = 0 \\ v(t+T_H) - v^G(t) = 0 \\ a(t+T_H) - a^G(t) = 0 \\ e(t+T_H) - e^G(t) = 0 \end{cases}$$

- the *partial-position* constraint, i.e.

$$p(t+T_H) - p^G(t) = 0 \quad \text{(Equation 9)}$$

- the *partial-position/velocity* constraint, i.e.

$$\begin{cases} p(t+T_H) - p^G(t) = 0 \\ v(t+T_H) - v^G(t) = 0 \end{cases} \quad \text{(Equation 10)}$$

Second, in the *full-state* constraint, the desired values of the constrained states are chosen. It is in general easy to set positional constraints (e.g. positions of via-points) and sometimes possible to set velocity constraints (e.g. stationary via-points). For the other states, two conservative methods were considered:

- a *zero-value method* (the states go to 0), e.g. when only the position goal is known

$$\begin{cases} p(t + T_H) - p^G(t) = 0 \\ \quad v(t + T_H) = 0 \\ \quad a(t + T_H) = 0 \\ \quad e(t + T_H) = 0 \end{cases} \quad \text{(Equation 11)}$$

or when the position and velocity goals are known

$$\begin{cases} p(t + T_H) - p^G(t) = 0 \\ v(t + T_H) - v^G(t) = 0 \\ \quad a(t + T_H) = 0 \\ \quad e(t + T_H) = 0 \end{cases} \quad \text{(Equation 12)}$$

- a *current-value method* (the states keep their current value), e.g. when only the position goal is known

$$\begin{cases} p(t + T_H) - p^G(t) = 0 \\ v(t + T_H) - v(t) = 0 \\ a(t + T_H) - a(t) = 0 \\ e(t + T_H) - e(t) = 0 \end{cases} \quad \text{(Equation 13)}$$

or when the position and velocity goals are known

$$\begin{cases} p(t + T_H) - p^G(t) = 0 \\ v(t + T_H) - v^G(t) = 0 \\ a(t + T_H) - a(t) = 0 \\ e(t + T_H) - e(t) = 0 \end{cases} \quad \text{(Equation 14)}$$

This description of boundary constraints applied to the via-points (goal states 1 to $N - 1$; Equation 6). The target (goal state $N$; Equation 6) was considered as stationary and always obeyed to Equation 11.

In summary, a task representation was defined by a series of goal, a goal setting policy (absolute or relative), a boundary constraint (full-state, partial-position, partial-position/velocity) and a boundary method for the full-state constraint (zero- or current-value). It should be noticed that the full-state constraint defined by Equation 11 or Equation 13 is not similar to the partial-position constraint (Equation 9). In the latter case, the values of velocity, activation, and excitation are constrained by optimization. For a similar reason, the full-state constraint defined by Equation 12 or Equation 14 is not similar to the partial-position/velocity constraint (Equation 10).

The apparent complexity in the construction of task representations is due to the fact that there are many implementation details and we deem it necessary to evaluate their contribution to the functioning of the model.

*Noise*

In the proposed framework, variability is modeled by the presence of white noise sources. We assumed that every process is possibly corrupted by noise and noise can be described by the standard deviation ($\sigma$) of a Gaussian variable. The dynamics (subscript *m* for motor; Equation 2) and observation (subscript *s* for sensory; Equation 5) noises were described in terms of signal-independent (SIN) and signal-dependent (SDN) terms ($SIN_m$, $\sigma^\xi$; $SDN_m$, $\sigma^\varepsilon$; $SIN_s$, $\sigma^\omega$; $SDN_s$, $\sigma^\epsilon$; Guigon et al., 2008b; Todorov, 2005). Other sources of noise were considered for:

- $T_H$, i.e. initially and at every $T_G$, the nominal receding horizon $T_H$ became $(1 + \sigma^\zeta \zeta)T_H$, where $\zeta$ is a realization of a standard normal variable and $\sigma^\zeta$ a parameter;

- $T_G$, i.e. at every $T_G$, the nominal goal time $T_G$ became $(1 + \sigma^\gamma \gamma)T_G$, where $\gamma$ is a realization of a standard normal variable and $\sigma^\gamma$ a parameter;

- $p^G$, i.e. initially the nominal position goal $p^G$ became $(1 + \sigma^\chi \chi)p^G$, where $\chi$ is a realization of a standard normal variable and $\sigma^\chi$ a parameter.

The dynamics and observation noises were termed execution noises as they act continuously during the movement. The other noises were termed planning noises as they act before the movement and possibly discretely during the movement (at each $T_G$).

*Parameters*

Parameters were $m = 1$ kg, $\tau = 0.05$ s, $T_H = 0.28$ s, $T_G = 0.13$ s, and **O** was the 4×4 identity matrix. For calculating the Kalman gain, the ratio $\sigma^\xi/\sigma^\omega$ was 0.001.

*Solutions*

The optimal feedback control policy and the Kalman gain were calculated analytically as described in Guigon et al. (2008b), and simulated numerically with the time step $\delta = 0.001$ s. A complete mathematical background is given in the Online Supplemental Material.

*Two-dimensional inertial point*

To simulate drawing movements, independent inertial points moving along perpendicular directions in the plane were considered (same formulation, same parameters).

*Data analysis*

Simulations produced time series of position that were analyzed to determine movement characteristics. Movement duration was defined by the time between the beginning of the simulation and the end of the movement obtained by a velocity threshold (0.05 m/s). Endpoint variability was defined as the standard deviation of the position at the end of the movement. Timing variability was defined as the standard deviation of movement duration.

## Fastest point-to-point movements: smoothness and isochrony

The fastest point-to-point movement was obtained when, starting from initial state $\boldsymbol{x}_0 = [0\ 0\ 0\ 0]$, a stationary target goal $\boldsymbol{x}^G = [p\ 0\ 0\ 0]$, ($p$ is the prescribed movement amplitude; full-state constraint; Equation 11) was set at time $t = 0$ and pursued at horizon $T_H$ (Figure 2A, dotted lines). The resulting trajectory was regular (Figure 2A) with a bell-shaped velocity profile (Figure 1B). As movement amplitude increased, peak velocity increased and movement duration remained constant (Figure 2B). These movements correspond to a unique class of smooth movements with biphasic acceleration profiles (Figure 2B, inset). They were close to minimum-jerk movements, with a jerk ratio of 1.5 (ratio between integrated jerk and jerk of the corresponding minimum-jerk trajectory). Note that the goal time $T_G$ played no role in these simulations since there were no via-points.

Two points should be noticed. First, by construction a movement obtained by control with a receding horizon never really stops in the sense that there always remains the same time to complete a movement irrespective of the remaining distance to the goal. In practice, movement duration can be properly defined by setting a threshold below which velocity fluctuations are negligible. The same consideration also applies to experimental recordings. Second, humans can produce movements that are faster than those described here (~0.4 s vs ~0.2 s in Hoffman & Strick, 1986). Much faster displacements (~0.2 s) can be obtained in the model in the framework of rhythmic movements (see Figure 12). The movements described by Hoffman and Strick (1986) have large overshoots and terminal oscillations, and could be considered as aborted rhythmic movements.

*Figure 2*. Simulation of the fastest point-to-point movements. **A**. Position profiles. Amplitude of 0.1 m (*black*), 0.2 m (*red*), 0.3 m (*green*), 0.4 m (*blue*). Dotted lines show task representation. **B**. Velocity profiles. Inset: acceleration profiles.

**Slow point-to-point movements: segmentation**

A slow movement of mean speed $s$ was obtained with the full-state constraint, either the zero-value (Equation 12) or the current-value (Equation 14) method, and the absolute goal setting policy (Equation 7). The time series of goal states was built according to rules 1 and 4: goal states were chosen along the desired spatial path of the movement and updated with period $T_G$. Starting from initial state $x_0 = [0\ 0\ 0\ 0]$, the first goal state was $x_1^G = [sT_G\ s\ 0\ 0]$, and the following goal states were given by Equation 7 with $\alpha_k = [sT_G\ 0\ 0\ 0]$. Accordingly, the position constraint was a staircase temporal signal of base $T_G$ and height $sT_G$ (dotted lines; Figure 3A), and the velocity constraint a constant signal of size *s* (dotted lines; Figure 3B). The simulated position and velocity profiles are shown in Figure 3 (plain lines) for 4 mean speeds. We showed previously that this model provides an accurate account of properties of slow movements (Guigon et al., 2019). Note that these movements can be terminated by setting a stationary target goal (Equation 6).

The relative goal setting policy (Equation 8) lead to slow movements which did not comply with the required mean speed. The partial-position/velocity constraint (Equation 10) lead to slow movements with smaller velocity fluctuations. As the value of $T_H$ was derived from the size of these fluctuations (Guigon et al., 2019), it raises the possibility for a lower value of $T_H$. Yet an independent cross-check of the value of $T_H$ was obtained in the study of drawing movements (see below).

*Figure 3*. Simulation of slow point-to-point movements. **A**. Position profiles. Mean speed of 0.025 m/s (*black*), 0.05 m/s (*red*), 0.075 m/s (*green*), 0.1 m/s (*blue*). Simulations with full-state constraint, zero-value method, and absolute goal setting policy. **B**. Velocity profiles. Dotted lines show task representation in position and velocity.

## Point-to-point movements of intermediate duration

Point-to-point movements that are neither very fast nor very slow have been ubiquitously reported in the literature. For instance, the mean movement speed in the original data of Fitts (1954) was in the range 10-100 cm/s (amplitude 5-40 cm, duration 0.18-0.73 s). The velocity profile of such movements is more or less irregular and asymmetric with one or more peaks (Darling, Cole, & Abbs, 1988; Morasso, Mussa-Ivaldi, & Ruggiero, 1983; Salmond et al., 2017). Following the proposed principles, the duration of a movement is related to the number and position of via-points used to define it. There is a priori an infinite number of ways to set via-points. In keeping with the study of slow movements, we considered movements built using via-points specified by a fraction of movement amplitude (rules 1, 2 and 4). For a movement of amplitude *A* and fraction *f* (in [0;1]), we took $\alpha_k = fA$ to define the position of the kth via-point (Equation 7 or Equation 8). A via-point whose position exceeded *A* was set at *A*, and considered as a target (Equation 6). Simulations were run for one amplitude and 10 fractions (between 0.3 and 1), with full-state constraint and zero-value method (Equation 11), for absolute (Equation 7; Figure 4A) and relative (Equation 8; Figure 4D) goal setting policy.

For the absolute goal setting policy, the via-points were equally distributed along the movements (Figure 4A), and the trajectories were regular (Figure 4B) with velocity profiles that became more irregular as movement duration increased (Figure 4C). The fastest movement was identical to those described in Figure 2 and the slowest movement was similar to those described in Figure 3. For the relative goal setting policy, the distribution of via-points along the movements was time-dependent (in fact, speed-dependent; Figure 4D), and the trajectories were as described for the absolute goal setting policy (Figure 4E,F). The main quantitative difference between the two goal setting policies is observed on the time to peak velocity which scales with duration in one case (Figure 4C), but not in the other (Figure 4F). This difference will be important in the study of scaling laws and Fitts' laws (see below).

*Figure 4*. Simulation of movements of intermediate duration. **A**. Position of via-points (*plain square*) for 10 movements of the same amplitude (*diamond*: start point; *circle*: target) with the absolute goal setting policy. **B**. Trajectories of the movements described in **A**. **C**. Velocity profiles of the movements described in **A**. **D**. Same as **A** the relative goal setting policy. **E**. Trajectories of the movements described in **D**. **F**. Velocity profiles of the movements described in **D**.

## Drawing movements: isochrony and power laws

A shape was described by a sequence of appropriately placed via-points updated at period $T_G$ (rules 2 and 4). For instance, four points at the vertex of a square presented sequentially at interval $T_G$ generated a circle figure (partial-position constraint, Equation 9; absolute goal setting policy, Equation 7; Figure 5A). The corresponding times serie of positional goals is shown in Figure 5B. No velocity goals were imposed at the via-points. Tangential velocity was inversely correlated with curvature and cycle duration was constant (Figure 5E). A more precise circle drawing was obtained with an 8-points sequence (Figure 5C,D). Drawing had similar scaling properties, but was slower (Figure 5E). Cycle duration increased linearly with the number of via-points (Figure 5E, inset).

For the same via-points, the partial-position/velocity (Equation 10) and the full-state constraints (Equations 11-14) lead to polygons rather than circles. The relative goal setting policy (Equation 8) was not appropriate to obtain a specific geometric shape.

*Figure 5.* Drawing circles. **A**. A 10-cm radius circle defined by 4 via-points (*colored squares*; color code from left to right on the colorbar), 30 turns, counterclockwise rotation. Mean cycle duration was 0.52 s. Calibration is 10 cm. **B**. Time series of positional goals in x (*thick*) and y (*thin*) for **A**. **C**. A 10-cm radius circle defined by 8 via-points (*colored diamonds*), 30 turns, counterclockwise rotation. Mean cycle duration was 1.04 s. **D**. Time series of positional goals in x (*thick*) and y (*thin*) for **C**. **E**. Relationship curvature/tangential velocity (*left axis, closed symbols*) and curvature/cycle duration (*right axis, open symbols*). Squares correspond to 4 via-points, diamond to 8 via-points. Dashed line from Viviani and McCollum (1983), fig. 3A. Inset: relationship between the number of via-points and cycle duration. Simulations with partial-position constraint, absolute goal setting policy, in the presence of $SDN_m$ ($\sigma^\varepsilon = 1$).

Studies of drawing curved shapes revealed a simple power law relationship

$$V(t) \propto C(t)^{-\beta}$$

between tangential velocity $V$ and curvature $C$ with $\beta \approx 1/3$ (Lacquaniti, Terzuolo, & Viviani, 1983). An ellipse defined by the four vertices of a rectangle (with an aspect ratio of 2) is shown in Figure 6A and complies with the one-third power law (Figure 6B).

Huh and Sejnowski (2015) showed that there exists a continuum of power laws related to the frequency content of shapes (see also Zago, Matic, Flash, Gomez-Marin, &

Lacquaniti, 2018). We built a via-point representation of 12 shapes (rules 2 and 4) corresponding to frequencies $\nu$ = 2/33, 2/5, 3/5, 2/3, 4/5, 4/3, 3/2, 2 (ellipse), 5/2, 3, 4, 5 according to the equation

$$log(r) = \mu sin(\nu\theta)$$

in polar coordinates $(r, \theta)$, $\mu$ is a parameter (chosen to obtain the shapes shown in Huh and Sejnowski, fig. 4). The exponent $\beta$ of the power law varied with frequency with a trend close to that reported experimentally (Figure 6C). The one-third power law was observed only for ellipses. A parametric analysis of these results shows that the chosen values of $T_H$ and $T_G$ provide the best account of experimental data over the range [0.18-0.48] for $T_H$ and [0.13-0.23] for $T_G$ (Appendix A, Figure A1).

*Figure 6.* Power laws. **A**. Ellipse defined by 4 via-points, 40 turns, counterclockwise rotation, arbitrary dimension. Same conventions as in Figure 5A. **B**. Instantaneous relationship between tangential velocity and curvature for the ellipse in **A**. $\beta = 0.328$ ($R^2 = 0.98$). **C**. Relationship between frequency content $\nu$ and power law exponent $\beta$ for 12 shapes. Dotted line from Huh and Sejnowski (2015), dashed line from Zago et al. (2018). Horizontal and vertical plain lines indicate the case of an ellipse ($\nu = 2$, $\beta \approx 1/3$). Simulations with partial-position constraint, absolute goal setting policy, in the presence of $SDN_m$ ($\sigma^\varepsilon = 1$).

## Handwriting

The word "flow" was used for comparison with Huys et al. (2014). It was described as a time series of 21 via-points (Figure 7A): the letter *f* was represented by two symmetrical rectangular triangles (*dark blue*), the letter *l* by an isocele triangle (*light blue*), the letter *o* by a smaller triangle (*green*) and the letter *w* by a kind of *w* (*yellow to red*). The methods were as described for drawing. The writing was smooth (Figure 7B), made of strokes (Figure 7C), and isochronous (by construction), consistent with experimental observations (Denier van der Gon, & Thuring, 1965; Hollerbach, 1981; Viviani & Terzuolo, 1982). This simulation is provided here for illustrative purpose. The via-points were chosen by hand. No general recipe for building word representations was searched for. In fact, there are no specific characteristics of handwriting that could be used to challenge the theory.

*Figure 7*. Writing the word "flow". **A**. Via-point representation of the word. The via-points are shown as nodes of a graph connected by colored edges. The order is indicated by the color gradient. **B**. Written word. The colors correspond to the via-point representation. **C**. Velocity profile. No spatial unit is necessary since writing is isochronous. Simulation with partial-position constraint, absolute goal setting policy, in the absence of noise.

## Scaling laws

Scaling laws correspond to ubiquitous relationships between movement characteristics (e.g. amplitude and duration; Fitts, 1954; Gordon, Ghilardi, Cooper, & Ghez, 1994). For instance, the fastest point-to-point movements of Figure 2 have constant duration and time to peak velocity, and linearly increasing peak velocity and peak acceleration with amplitude (Figure 8A). We call this pattern an *isochronous scaling law*. To analyze the diversity of scaling laws, we considered movements built as described for movements of intermediate duration (Figure 4A,D). Simulations were run for 4 amplitudes and 10 fractions (between 0.5 and 1), with full-state constraint and zero-value method (Equation 11), for absolute (Equation 7; Figure 8B) and relative (Equation 8; Figure 8C) goal setting policy. For $f < 1$, there was one via-point for absolute goal setting policy (Appendix B, Figure B1A), and one or two via-points for relative goal setting policy (Figure B1B). We observed that each fraction $f$ corresponded to a specific isochronous scaling law (one color; Figure 8B,C). For each amplitude, movement duration increased and peak velocity/acceleration decreased as $f$ decreased. As already observed (Figure 4C,F), the two goal setting policies differed on the time to peak velocity: it scaled with $f$ for absolute goal setting policy and was constant for relative goal setting policy (Figure 8B,C). The scaling laws remained similar for the current-value method (Equation 13).

> *Figure 8*. Scaling laws. **A**. Scaling laws for the fastest point-point movements (data from Figure 2). (*circle*) movement duration, left side axis; (*square*) time to peak velocity, right side axis. **B**. A family of isochronous scaling law obtained with full-state constraint, zero-value method and absolute goal setting policy. Each point corresponds to a movement of a given amplitude and a specific position of via-points. Each color corresponds to four movements (amplitudes 0.1, 0.2, 0.3, 0.4 m) with a specific fraction (10 fractions between 0.5 and 1) of amplitude between the via-points. For instance a fraction of 0.6 means that the four movements have one via-point at 0.06, 0.12, 0.18 and 0.24 m, respectively. **C**. Same as **B** with relative goal setting policy.

From a family of isochronous scaling laws (Figure 8B,C), different types of scaling strategies could be obtained (Figure 9):

- via-points at a decreasing fraction of amplitude with absolute goal setting policy gave movements with increasing duration, time to peak velocity, peak velocity, and peak acceleration (Figure 9A; Gordon et al., 1994);

- via-points at a fixed distance with absolute goal setting policy gave movements with increasing duration, peak velocity and time to peak velocity, but constant peak acceleration (Figure 8B; non-dominant strategy in Sainburg & Schaeffer, 2004);

- via-points at a decreasing fraction of amplitude with relative goal setting policy gave movements with increasing duration, peak velocity and peak acceleration, but constant time to peak velocity (Figure 9C; dominant strategy in Sainburg & Schaeffer, 2004).

The model can thus account for the fact that the kinematic landmarks of a movement do not systematically scale with changes in movement duration (Baraduc, Thobois, Gan, Broussolle, & Desmurget, 2013; Sainburg & Schaeffer, 2004; Sarlegna, Blouin, Bresciani, Bourdin, Vercher, & Gauthier, 2003; Torres & Andersen, 2006; Worringham, 1991). The model does not provide a rationale for the different strategies, but this issue will become clearer in the framework of Fitts' law (see below).

*Figure 9*. Three scaling strategies. **A**. Absolute goal setting policy, via-points at fraction 0.9, 0.8, 0.7 and 0.6, for amplitude 0.1, 0.2, 0.3, 0.4 m, respectively. **B**. Absolute goal setting policy, via-points at 0.11 m for all amplitudes. **C**. Relative goal setting policy, via-points at fraction 0.9, 0.8, 0.7 and 0.6, respectively. Same conventions as in Figure 8. Below each scaling strategie, the position of the via-points for each amplitude is shown (same format as in Figure 4A,D).

## Discrete and rhythmic movements

There exists a classical distinction between discrete and rhythmic movements (Guiard, 1993; Hogan & Sternad, 2007). An open question is whether the model can account for this distinction. To produce a rhythmic behavior, we considered a sequence of two alternating position goals (distance 0.2 m) at a fixed frequency (corresponding to an internally generated periodic signal) as a task representation (rules 1 and 3), with partial-position constraint (Equation 9). At 1.5 Hz, position, velocity and acceleration traces were almost sinusoidal (Figure 10A). There were no dwell periods (i.e. periods of near-zero velocity) and harmonicity (defined as the ratio between minimum and maximum acceleration; Guiard, 1993) was close to one. At 0.7 Hz, kinematics was much more irregular with consistent dwell periods and near zero harmonicity (Figure 10B). Dwell time and amplitude decreased with frequency (Figure 10C), and harmonicity increased with frequency (Figure 10D). Qualitatively, nonzero dwell times were found for frequencies below 1 Hz, i.e. for unidirectional displacements longer than 0.5 s, consistent with Sternad, Marino, Charles, Duarte, Dipietro, & Hogan (2013). The corresponding movements are considered as discrete (Hogan & Sternad, 2007). We note that movements of such durations ($> 0.5$ s) could be produced using via-points (see Figure 8; rule 5) and would not encompass a dwell time. Using either the partial-position/velocity or the full-state constraint lead to qualitatively similar results at high frequencies and a shift to nonzero dwell times around 2 Hz. These results do not depend on the goal time $T_G$ since the goals were updated at imposed frequencies.

These results show that the distinction between discrete and rhythmic movements could correspond to a change in behavior (appearance of dwell periods) with required movement frequency of one and the same control process. Further analysis of rhythmic movements is given in Appendix C.

*Figure 10.* Rhythmic movements. **A**. (*top*) Position and acceleration (*bottom*) Velocity. Vertical dashed and dotted lines delimit dwell periods (velocity threshold at 0.01 m/s). Frequency was 1.5 Hz. Dwell time was 0.015 s. Harmonicity was 0.67. **B**. Same as **A** for a frequency of 0.7 Hz. Dwell time was 0.172 s. Harmonicity was 0.28. **C**. Mean dwell time as a function of frequency. Inset: mean movement amplitude as a function of frequency. (*gray circle*) 1.5 Hz; (*black circle*) 0.7 Hz. **D**. Mean harmonicity as a function of frequency. Simulations with partial-position constraint, in the presence of $SDN_m$ ($\sigma^\varepsilon = 2$). Simulation duration was 120 s.

## Fitts' law

In the presence of noise, scaling laws (see above) define a tripartite relationship between movement amplitude, duration and endpoint variability. We explored this relationship for discrete and rhythmic (reciprocal) movements.

*Discrete Fitts' law*

We simulated series of movements of different amplitudes and with via-points at different positions corresponding to the scaling strategies of Figure 8B (absolute goal setting policy) and 8C (relative goal setting policy) in the presence of $SDN_m$ (see below for the influence of noise on Fitts' law; Figure 16). We measured actual movement amplitude ($A$), duration ($MT$, to keep a traditional notation) and endpoint variability ($W$), and we analyzed the relationship between the effective index of difficulty $ID = log_2(2A/W)$ and $MT$. As expected from Fitts' law, *MT* was a linear function of *ID* (Figure 11A,B). We note that the actual range of *ID* is arbitrary: changing the level of noise or the definition of endpoint variability would shift the range of *ID* with no effect on *MT* (this remark applies to all the simulations involving a calculus of *ID*). Time to peak velocity and time after peak velocity increased with *ID* for absolute goal setting policy (Figure 11C) while time to peak velocity remained constant for relative goal setting policy (Figure 11D). These two strategies were observed in the study of discrete Fitts' law (Poletti, Sleimen-Malkoun, Temprado, & Lemaire, 2015; Sleimen-Malkoun, Temprado, & Berton, 2013; Temprado, Sleimen-Malkoun, Lemaire, Rey-Robert, Retornaz, & Berton, 2013). Note that, for each amplitude, peak velocity decreased with *ID* for the relative goal setting policy (Figure 11B, inset; MacKenzie, Marteniuk, Dugas, Liske, & Eckmeier, 1987). Fitts' law was observed with the zero-value method, but not with the current-value method.

*Figure 11*. Discrete Fitts' law. **A**. Movement duration as a function of *ID*. 5 amplitudes (0.1-0.5 m), via-points at 5 equi-spaced distances (0.11, 0.132, 0.155, 0.178, 0.2 m). Simulations with full-state constraint, zero-value method, absolute goal setting policy. **B**. Movement duration as a function of *ID*. 5 amplitudes (0.1-0.5 m), via-points at 5 fractions (0.65, 0.725, 0.8, 0.875, 0.95). Simulations with full-state constraint, zero-value method, relative goal setting policy. Inset: peak velocity as a function of *ID*. (*diamond*) 0.1 m; (*up triangle*) 0.2 m; (*down triangle*) 0.3 m; (*left triangle*) 0.4 m; (*right triangle*) 0.5 m. **C**. Time to peak (*black*) and time after peak (*gray*) velocity as a function of function of *ID* for data in **A**. **D**. Time to peak (*black*) and time after peak (*gray*) velocity as a function of function of *ID* for data in **B**. Simulations in the presence of $SDN_m$ ($\sigma^\varepsilon = 1$). Variability calculated over 500 trials per condition.

*Rhythmical (reciprocal) Fitts' law*

We simulated series of rhythmical movements of a given amplitude as explained above (Figure 10) for frequencies in the range 0.7-2.5 Hz. As movement amplitude decreases with frequency (Figure 10C, inset), compensatory changes in movement amplitude with frequency were used to obtain the desired amplitude (Figure 12B, inset). For the given amplitude, movement duration and time to peak velocity increased linearly with *ID* for *ID* below 4 (Figure 12A). Dwell time increased with *ID* (Figure 12B) and harmonicity decreased for *ID* below 4 (Figure 12C). These observations are consistent with experimental results for low *ID*s (Boyle & Shea, 2013; Buchanan, 2013). The behavior changed above *ID* = 4 (low frequencies) consistent with a transition between rhythmic and discrete movements (Huys, Fernandez, Bootsma, & Jirsa, 2010). The data for *ID* > 4 are shown, but are wrong in the sense that a different mechanism is involved at higher IDs (lower frequencies; see above, Figure 11B,D).

The rhythmical movements obeyed to Fitts' law, i.e. a univocal relationship was found between *ID* and movement duration for different amplitudes (Figure 13A). Peak velocity decreased with *ID* as observed experimentally (Figure 13B; Boyle & Shea, 2013). A rhythmical Fitts' law was not observed for partial-position/velocity and full-state constraints.

*Figure 12*. Rhythmical Fitts' law (single amplitude). **A**. Movement duration (*circle*) and time to peak velocity (*square*) as a function of *ID*. Amplitude was 0.2 m. **B**. Dwell time. Inset: compensatory changes in amplitude as a function of frequency. Dotted line corresponds to 0.2 m (desired amplitude). **C**. Harmonicity. (*gray lines*) data from Buchanan (2013). 15 frequencies in the range 0.7-2.5 Hz. Simulations with partial-position constraint, in the presence of $SDN_m$ ($\sigma^\varepsilon = 2$). Variability calculated over 500 trials per condition.

*Figure 13*. Rhythmical Fitts' law (multiple amplitudes). **A**. Movement duration as a function of *ID* for 3 amplitudes: (*red*) 0.1 m; (*black*) 0.2 m; (*green*) 0.3 m. **B**. Peak velocity as a function of *ID*. Same parameters as in Figure 12.

## Timing

Movements can be made toward spatial targets, but also toward temporal events (e.g. the beat of a metronome; Howarth, Beggs, & Bowden, 1971). There is evidence that movements directed at spatial and temporal goals are subserved by different processes (Howarth et al., 1971; Huys et al., 2010). For the model, this means that the task representation used to produce rhythmic movements (see above) is not appropriate to explain how movements get synchronized with a metronome. An attempt to address this issue is the following. We observe that the kinematics of a movement toward a temporal goal have peculiar features, i.e. its velocity profile is asymmetrical and peak velocity occurs late in the movement (Cos, Girard, & Guigon, 2015; Craig, Pepping, & Grealy, 2005; Port, Lee, Dassonville, & Georgopoulos, 1997; Rieger, 2007; Walter & Rieger, 2012). Furthermore time to peak velocity increases all the more so that the period of the metronome increases (Cos et al., 2015). We assume that this property is specific to timing as it has never been reported in other motor conditions. We searched for a task representation of timing in terms of via-points. The size of peak velocities depends on the distance between via-points, thus to obtain larger velocity peaks later in the movement, it is sufficient to consider nearby via-points at the begin of the movement and more distant via-points later. Accordingly, the distance between the via-points $n-1$ and $n$ was chosen to be $nfA$ where $A$ is the movement amplitude and $f$ a parameter in [0,1]. Simulated velocity profiles are shown in Figure 14A. Time to peak velocity increased with movement duration and occurred later in proportion of duration (Figure 14B). Variability in movement duration increased with movement duration in the presence of noise on the goal time (Figure 14A, inset). $SDN_m$ was not sufficient to obtain this kind of variability.

*Figure 14*. Simulation of movements with late peak velocity. **A**. Velocity profiles. See Text for explanation. Colors from blue to red: $f = 0.025, 0.05, 0.075, 0.1, 0.125, 0.15, 0.2, 0.3$. Movement amplitude was 0.2 m. Inset: standard deviation of movement duration as a function of movement duration. **B**. Relationship between movement duration and time to peak velocity in s (*left axis, open squares*) and % of movement duration (*right axis, closed squares*). Simulations with full-state constraint, zero-value method, relative goal setting policy, in the presence of $SDN_m$ ($\sigma^\varepsilon = 2$) and noise on the goal time ($\sigma^\gamma = 0.005$). Variability calculated over 500 trials per condition.

We assessed the relationship between amplitude, duration and variability for timing movements as we did for the study of Fitts' law. Movement duration increased with the index of difficulty, but the slope was steeper than for a Fitts' task (Figure 15A; Howarth et al., 1971; Huys et al., 2010). Spatial variability increased linearly with average velocity (Figure 15B; Schmidt, Zelaznik, Hawkins, Franck, & Quinn, 1979; Wright & Meyer, 1983). Temporal variability increased linearly with movement duration (Figure 15C; Bongers, Fernandez, & Bootsma, 2009). These results show that the model can account for specific characteristics of speed/accuracy tradeoff in a timing task. We note that the model is unable to account for the presence of pauses that precede movements toward temporal events in synchronization tasks (Cos et al., 2015; Donnet, Bartolo, Fernandes, Cunha, Prado, & Merchant, 2014; Hove & Keller, 2010).

The robustness of these results is the following. The results remained qualitatively similar under the current-value method, for partial position-velocity and partial-position constraint, in the absence of $SDN_m$. The structure of temporal variability (Figure 15C) disappeared in the absence of temporal noise. The results were lost with absolute goal setting, in the presence of $SDN_s$ and with noise only on the receding horizon.

*Figure 15*. Tripartite relationship between amplitude, duration and variability for timing movements. **A**. Movement duration as a function *ID*. Three movement amplitudes (0.2, 0.3, 0.4 m) and 10 fractions (*f* between 0.02 and 0.2). Gray lines from Howarth et al. (1971), fig. 3: (*dashed*) data from Fitts (1954); (*solid*) timing task. **B**. Spatial variability as a function of average velocity. **C**. Temporal variability as a function of movement duration. Simulations with full-state constraint, zero-value method, relative goal setting policy, in the presence of $SDN_m$ ($\sigma^\varepsilon = 2$) and noise on the goal time ($\sigma^\gamma = 0.005$). Variability calculated over 500 trials per condition (amplitude, fraction).

## Structure of variability and the nature of noise

In the preceding simulations, we used sources of white Gaussian noise to explain variability. We address here the role of the different sources.

We have shown that discrete Fitts' law can be explained by the presence of $SDN_m$ (Figure 11). We assessed the basic tripartite relationship between amplitude, duration and variability for the different types of noise (except $SIN_m$ which has a deleterious effect on control). The simulations confirmed the proper role of $SDN_m$ (Figure 16, *black*; see Figure 11A) and showed that variability produced by the other sources of noise did not comply to Fitts' law, i.e. there was no univocal relationship between movement duration and ID (Figure 16, *red, green, blue, purple, yellow*).

*Figure 16*. Simulation of discrete Fitts' law for different sources of noise. (*black*) $SDN_m$ ($\sigma^\varepsilon = 2$); (*red*) $SIN_s$ ($\sigma^\omega = 0.0001$); (*green*) $SDN_s$ ($\sigma^\epsilon = 0.5$); (*blue*) $T_H$ ($\sigma^\zeta = 0.2$); (*purple*) $T_G$ ($\sigma^\gamma = 0.2$); (*yellow*) $x^G$ ($\sigma^\chi = 0.01$). Noise standard deviations were chosen to obtain similar ranges of *ID*, except for noise on $T_G$ which cannot produce a larger variability. Only the type of noise indicated was present. 3 amplitudes (0.2, 0.25, 0.3 m). Variability calculated over 500 trials per condition.

When participants repeatedly produce movements of a given amplitude, movement duration (*MT*) covaries with peak velocity (*PV*), i.e. $PV \propto MT^{-q}$ with $q \approx 0.8$ (van Beers, Haggard, & Wolpert, 2004; see also Messier & Kalaska, 1999). This property was found in the model for $SDN_m$, noise on $T_H$ and noise on $T_G$ (Figures 17A,B,C), although the velocity profiles were peculiar for the two latter noises (Figure 17A, inset; 17B).

*Figure 17*. Variability of velocity profiles. **A**. Velocity profiles (*gray lines*) and peak velocity (*black circle*) for $SDN_m$ ($\sigma^\varepsilon = 2$). 100 trials are shown. Inset: same for $T_H$ (*blue*) and $T_G$ (*purple*) noise. **B**. Relationship between peak velocity and movement duration for the data in **A**. 500 trials are shown. **C**. Exponent $q$ of the relationship between peak velocity and movement duration vs $R^2$ of the relationship. Vertical dashed line indicates $q = 0.8$ (expected value; van Beers et al., 2004). (*black*) $SDN_m$ ($\sigma^\varepsilon = 2$); (*red*) $SIN_s$ ($\sigma^\omega = 0.00005$); (*green*) $SDN_s$ ($\sigma^\epsilon = 0.25$); (*blue*) $T_H$ ($\sigma^\zeta = 0.18$); (*purple*) $T_G$ ($\sigma^\gamma = 0.2$); (*yellow*) $x^G$ ($\sigma^\chi = 0.004$). Standard deviations of noise were chosen to obtain an $ID \approx 7$, except for $T_G$. Simulations with full-state constraint, zero-value method, absolute goal setting policy. Amplitude 0.1 m, via-point at 0.07 m, chosen to obtain a movement duration around 0.4 s as in van Beers et al. (2004). Variability calculated over 500 trials.

We assessed basic aspects of correlation, spatial and temporal variability across repetitions of a movement of a given amplitude (Figure 18): 1. correlation between position at kinematic landmarks and endpoint position (Heath, Westwood, & Binsted, 2004; Figure 18A); 2. correlation between position at normalized time and endpoint position (Heath, Neely, & Krigolson, 2008; Figure 18B); 3. correlation between the value of kinematic landmarks (Messier & Kalaska, 1999; Figure 18C); 4. positional standard deviation at normalized time (Liu & Todorov, 2007; Figure 18D); 5. positional standard deviation of kinematic landmarks (Khan, Elliot, Coull, Chua, & Lyons, 2002; Figure 18E); 6. temporal standard deviation of kinematic landmarks (Valdez & Amazeen, 2008; Figure 18F). Each panel in Figure 18 shows simulations with 6 types of noise (*colored lines*, *circles*) and experimental data (*gray lines*, *triangles*) which are characteristic of properties reported in the literature. We observed again that the presence of $SDN_m$ (*black lines* compared to *gray lines*) provides a good quantitative account of the structure of variability of point-to-point movements. The presence of the other sources of noise led to contrasted results in particular as far as correlations were concerned (Figure 18A,B,C). Interestingly, both execution and planning noises can account for classical observations on the time course of positional variance (Figure 18D; Krüger, Eggert, & Straube, 2011; Krüger, Straube, & Eggert, 2017; Liu & Todorov, 2007; Mosier, Scheidt, Acosta, & Mussa-Ivaldi, 2005; Osu, Morishige, Nakanishi, Miyamoto, & Kawato, 2015; Selen, Beek, & van Dieën, 2006) which is generally thought to arise from execution noise (Guigon et al., 2008b; Todorov & Jordan, 2002).

*Figure 18*. Variability of kinematic landmarks. **A**. Correlation coefficient between position at peak acceleration (PA), peak velocity (PV) and peak deceleration (PD) and endpoint position. The green curve is superimposed on the red one. Data from Heath et al. (2004), fig. 7. **B**. Correlation coefficient between spatial position at normalized times and endpoint position. Data from Heath et al. (2008), fig. 7. **C**. Mean correlation coefficient between kinematic markers (A, amplitude). Data from Messier and Kalaska (1999), fig. 11. **D**. Positional standard deviation at normalized times. Data from Liu and Todorov (2007),

fig. 1e. **E**. Positional standard deviation at kinematic markers in percent of movement amplitude. Data from Khan et al. (2002), fig. 5. **F**. Temporal standard deviation at kinematic markers in percent of movement amplitude. Data from Valdez and Amazeen (2008), fig. 8. (*gray, triangle*) experimental data; (*black*) $SDN_m$ ($\sigma^\varepsilon = 2$); (*red*) $SIN_s$ ($\sigma^\omega = 0.000075$); (*green*) $SDN_s$ ($\sigma^\epsilon = 0.4$); (*blue*) $T_H$ ($\sigma^\zeta = 0.25$); (*purple*) $T_G$ ($\sigma^\gamma = 0.2$); (*yellow*) $\boldsymbol{x}^G$ ($\sigma^\chi = 0.01$). Standard deviations of noise were chosen to obtain similar ranges of positional standard deviation (**D**), except for $T_G$. Simulations with full-state constraint, zero-value method, absolute goal setting. Amplitude 0.2 m, via-point at 0.18 m. Variability calculated over 500 trials.

# Discussion

We presented an ensemble of simulations of motor behaviors designed following two guidelines. First they all involved the same control process. Second they each contained a specific description of a task (task representation) in terms of a spatiotemporal sequence of goals. The sequence was in general naturally related to the task, sometimes not so naturally but intuitively related to the task (movements of intermediate duration), and in a case the construction was more speculative (timing). If all the results may not be equally convincing, the overall impression is that the proposed theory has a large explanatory power that extends across many aspects of motor control. Several aspects of this work (submovement, intermittency) have been comprehensively discussed previously (Guigon et al., 2019) and are not addressed here.

## Disclaimer

The present work should not be considered as a subservient extension of a dominant theory (computational theory of motor control; Scott, 2004; Todorov & Jordan, 2002; Todorov, 2004; Wolpert & Ghahramani, 2000), but as the outcome of a slow and in-depth progression in the field of motor control (Guigon, 2010, 2011; Guigon et al., 2019; Rigoux & Guigon, 2012). It acknowledges past and present debates on motor control (Ajemian & Hogan, 2010; Feldman & Levin, 1995; Friston, 2011; Mohan, Bhat, & Morasso, 2019; Turvey, 1977), recognizes the existence of divergent positions, and makes some attempts toward a reconciliation. Using a receding horizon renders the control process stationary (i.e. independent of time) and intrinsically flexible in agreement with the dynamical systems approach to motor control (Kelso, 1995). The receding horizon also provides a natural continuity between movement and posture (Guigon, 2010) as in the equilibrium-point theory (Feldman & Levin, 1995). Yet the proposed theory remains sensitive to numerous criticisms attached to the representational view of action (Warren, 2006). Instead of trying to argue in vain against these criticisms, we developed two points based on the position/force dichotomy and the structured nature of motor variability that illustrate the power of the computational approach over other modeling frameworks (see below **Debates on motor control**). It is not a manoeuvre to bypass a debate, but a way to focus on technical arguments that are amenable to experimental testing rather than conceptual and philosophical arguments.

## Time in motor control

A central difficulty of the computational approach to motor control is how time is processed during the course of an action. By construction, no movement is possible without choosing in advance a time horizon (Equation 1). If no other principle is available, the time horizon remains fixed along the course of the movement, leading to a nonstationary control policy and an absence of flexibility in time. A paradoxical aspect of the proposed theory is to address these difficulties by setting two rigidly defined time constants (receding horizon, update period of goals). Control with a receding horizon benefits from an automatic updating of the time to achieve goals, which is necessary when perturbations are encountered (Liu & Todorov, 2007; Pélisson, Prablanc, Goodale, & Jeannerod, 1986; Shadmehr & Mussa-Ivaldi, 1994). Although the initial time horizon is set to the receding horizon, actual movement duration is an emergent property of the interaction between the controller and the environment. Movement duration also depends on self-selected goals (via-points) updated at a fixed period. Simple rules for the choice of via-points produce realistic scaling laws, speed/accuracy tradeoffs, and synchronization movements. A time series of spatial goals is an appropriate proxy for the specification of movement duration, precision and timing. The theory provides a principled approach to time processing in motor control which is consistent with the existence of isochronous behaviors and the nonsmooth aspect of most movements expect the fastest ones.

## Isochrony

A core property of the model is the production of isochronous actions. The origin of isochrony is unknown, but is considered here as a native characteristic of the model that underlies all other emergent properties. This view is consistent with the idea that time is given priority over space in motor control (Ashton, 1976; Lashley, 1951). For instance, in speech production, consistent timing is essential for intelligibility of speech (Faulkner & Rosen, 1999) and is preserved in patients with Parkinson's disease at the expense of amplitudes and velocities of lip displacements (Caligiuri, 1987; Connor, Abbs, Cole, & Gracco, 1989; Walsh & Smith, 2012). In the production of rhythmic arm movements, Levy-Tzedek, Ben Tov, & Karniel (2011) have shown that the participants exerted a stronger control on movement frequency than on amplitude or velocity.

Isochrony is not a priori a unitary concept. It has been observed for fast, discrete movements (Bryan, 1892; Jeannerod, 1984) and for slower, continuous motor activities (drawing, handwriting; Denier van der Gon & Thuring, 1965; Lacquaniti, Terzuolo, & Viviani, 1983; Viviani & McCollum, 1983). In the former case, isochrony could be a direct consequence of control with a receding horizon. In the latter, it could be due to both the receding horizon and the representation of task by series of goals updated at a fixed frequency, i.e. isochronous actions are those that have the same number of via-points irrespective of their amplitude (Figure 5). The failure to keep being perfectly isochronous in drawing circle of increasing size (Bennequin, Fuchs, Berthoz, & Flash, 2009; Viviani & Cenzato, 1985; Viviani & Schneider, 1991) could be explained by a strategic increase in the number of via-points with the perimeter of the circle to avoid an excessive increase in energetic expenditure, peak velocity, or variability (speed/accuracy tradeoff). Interestingly, the principles that lead to isochrony also account for the existence of power laws (Figure 6; Huh & Sejnowski, 2015; Lacquaniti et al., 1983; Zago et al., 2018). Accordingly, the framework proposed by Bennequin et al. (2009) to explain isochrony and power laws in terms of multiple geometries might be unwarrantedly complex.

An open question is whether an optimality principle could account for isochrony. As noted previously, a cost of time (Harris & Wolpert, 2006; Hoff, 1994; Shadmehr et al., 2010) could explain the compensatory regulation of movement speed with amplitude to maintain movement duration approximately constant. Yet we would never obtain a strict isochrony since it is not a rational behavior to spend more cost to keep a constant duration. A possible normative account of isochrony can be derived in the framework of reward/effort-based optimal motor control (Rigoux & Guigon, 2012). We assume that the utility of producing an action of intensity $I$ (e.g. force, amplitude) and duration $T$ to obtain a reward $\rho$ is $J = \rho/T - \varepsilon I^2/T^2$ (which is a simplified version of the model developed in Rigoux & Guigon, 2012; $\varepsilon$ is a conversion factor) which defines a trade-off between the benefits and costs of an action. Maximum utility (with respect to $T$) is $J^* = \rho^2/4\varepsilon I^2$ and is obtained for $T^* = 2\varepsilon I^2/\rho$. We also assume that the reward $\rho$ can be interpreted as a "motivational investment", i.e. how hard we are ready to work for a given action, and we search for a function $\rho = \rho(I)$ which maximizes

$$\int \frac{J^*(I)}{\rho(I)} dI \ (Equation\ 15)$$

(ratio between utility and investment across the range of action intensities). Using the calculus of variations, the function $\rho$ should obey to the differential equation

$$I^2\dot{\rho}(I) - 2I\rho(I) = 0,$$

which gives $\rho(I) = KI^2$ where $K$ is a constant. The optimal utility is $J^* = K\rho/4$ and the optimal time is $T^* = 2/K$ which is constant. In this framework, isochrony is the outcome of an optimization process (Equation 15). Although it is interesting to have a normative explanation of isochrony, it is unclear whether this explanation is superior and more informative than a basic explanation based on the role of isochrony in the synchronization of multiple neural and behavioral events.

## Handwriting

Morasso and Mussa-Ivaldi (1982) distinguished muscle-oriented and space-oriented models of trajectory formation. In the former models, geometrical and mechanical (elastic) properties of muscles create parametric oscillatory dynamics that produce curved trajectories, e.g. drawings, letters (Hollerbach, 1981; Singer & Tishby, 1994). These models foreshadowed refined developments in the framework of the dynamical approach to motor control in which interactions on multiple time scales and between multiple dynamical patterns produce sequential, coordinated actions (e.g. handwriting; Huys et al., 2014; Perdikis, Huys, & Jirsa, 2011; for a different, but related work, see also Friston, Mattout, & Kilner, 2011). The latter space-oriented models are based on trajectory control in space either through combinations of motor primitives (strokes; Bullock, Grossberg, & Mannes, 1993; Edelman & Flash, 1987; Morasso & Mussa-Ivaldi, 1982) or through via-point representations (Gilet, Diard, & Bessière, 2011; Meulenbroek, Rosenbaum, Thomassen, Loukopoulos, & Vaughan, 1996;

Wada & Kawato, 1995; the present model). Most of these models produce realistic writing patterns, and many characteristic properties of handwriting such as isochrony and curvature/velocity relationship easily emerge. A more challenging issue is the absolute time scale of writing processes, e.g. the duration of individual strokes. Although it is difficult to pinpoint a single number (due to the heterogeneity of acquisition and data processing methods, in particular the filtering cutoff frequency), a realistic range is 0.1-0.3 s (Teulings & Stelmach, 1991). The value obtained here is ~0.19 s (16 velocity peaks in 3 s; Figure 6C) and is constant by construction. None of the models discussed above (nor other models) can account for the existence of a fixed duration of strokes because the processing time scale is in general determined by some free parameters (coefficients of a dynamics, stiffness, …). The strength of the present theory is to enable the coexistence of processing at a constant time scale on the one hand and full temporal flexibility on the other hand.

## Fitts' law

The fact that the model accounts for Fitts' law is not a surprise. The tripartite relationship between amplitude, duration and variability appears to be a ubiquitous consequence of control in the presence of multiplicative motor noise (Guigon et al., 2008a; Harris & Wolpert, 1998; Hoff & Arbib, 1992; Meyer, Abrams, Kornblum, Wright, & Smith, 1988; Qian et al., 2013; Rigoux & Guigon, 2012; Tanaka, Krakauer, & Qian, 2006). The main contribution of the theory is to explain 3 types of Fitts' law (discrete, rhythmic, temporal) and the kinematics of movements that obey to these laws. This corresponds to a major improvement in the computational description of Fitts' law since previous models accounted for a single type of Fitts' law and only in terms of unrealistic smooth movements of arbitrary duration (Guigon et al., 2008a; Harris & Wolpert, 1998; Jean & Berret, 2017).

# Motor variability

Variability is a central theme in the study of motor control that goes beyond Fitts' law. Common observations are that no two movements toward the same goal are exactly the same and variability calculated along repeated trajectories is not uniform (structured variability; Darling et al., 1988; Todorov & Jordan, 2002). Although most motor control theorists often cite Bernstein and agree with his claim that "the global structure of the movement remains the same, while the (details of the) movements never exactly repeat themselves" (citation from Meijer & Bruijn, 2007; original article of Bernstein in Feldman & Meijer, 1999), observations on motor variability and its structure have raised little interest and contributed little to theoretical constructs except in the computational framework (for an exception see Martin, Reimann, & Schöner, 2019). It could be revealing to see how variability emerges in other frameworks.

Admittedly, the methodology used for the study of variability has obvious limitations. First, noise is modeled as a white Gaussian stochastic process, although other types of noise exist. Second, part of the motor variability may not be uniquely related to the presence of noise (e.g. existence of trial-to-trial corrective processes; van Beers, Brenner, & Smeets, 2013) or to the presence of uncorrelated noise (e.g. existence of long-range correlations in series of motor actions; Slifkin & Eder, 2012). Yet, despite these limitations, consistent conclusions can be drawn. First, signal-dependent (multiplicative) motor noise is a major determinant of variability (Figures 16, 17, 18) consistent with previous experimental and theoretical results (Guigon et al., 2008b; Harris & Wolpert, 1998; Jones, Hamilton, & Wolpert, 2002; Meyer et al., 1988; Todorov, 2002; Todorov & Jordan, 2002). Second, planning noise contributes to movement variability, in particular for the production of temporal variability (Churchland, Afshar, & Shenoy, 2006; van Beers et al., 2004), and possibly for the emergence of speed/accuracy tradeoff (Al Borno, Vyas, Shenoy, & Delp, 2020).

## Discrete and rhythmic movements

There is strong evidence that discrete and rhythmic movements are not of the same nature, i.e. neither a rhythmic movement results from the concatenation of discrete segments nor a discrete movement can be identified to a truncated rhythmic movement (Guiard, 1993; Hogan & Sternad, 2007; van Mourik & Beek, 2004). Theoretical considerations on this distinction based on motor primitives (e.g. central pattern generators) have been reviewed by Degallier and Ijspeert (2010). The central theme was related to nonlinear dynamical equations whose solution corresponds to discrete or rhythmic patterns depending on parameters (Degallier, Righetti, Natale, Nori, Metta, & Ijspeert, 2008; de Rugy & Sternad, 2003; Huys, Studenka, Rheaume, Zelaznik, & Jirsa, 2008; Jirsa & Kelso, 2005; Ronsse, Sternad, & Lefèvre, 2009; Schaal, Kotosaka, & Sternad, 2000; Schöner, 1990). None of the described models can account for the changes in characteristics of rhythmic movements with frequency and the transition to discrete movements at lower frequencies (Guiard, 1993; Sternad et al., 2013).

An alternative approach was described by Biess, Nagurka, & Flash (2006) who defined discrete and rhythmic movements as solutions to optimal control problems with different boundary conditions. Our view of the difference between discrete and rhythmic movements is consistent with this idea. In this framework, rhythmicity is not the consequence of a continuous, autonomous dynamics, but of a discrete, periodic guidance. It should be noted that the transition between rhythmic and discrete movements observed around 1 Hz in the model does not correspond to a qualitative change in behavior (e.g. a bifurcation). A velocity threshold was used to detect "artificial" dwell periods, but in fact, the underlying velocity pattern changed with frequency along a continuum.

According to the theory, the computational specificity of rhythmic movements is related to the management of constraints, i.e. goals are defined only by positional constraints (Equation 9). Discrete movements require a broader set of constraints for the extra demands of termination, and accordingly, larger computational resources. This view is consistent with the observation that the neural substrate of rhythmic movements is a subset of the neural substrate of discrete movements (Schaal, Sternad, Osu, & Kawato, 2004).

## What makes the theory work?

The first principle (optimal feedback control) is responsible for the basic machinery of motor control (Todorov & Jordan, 2002) and the present model adds nothing specific to this point. The second principle (receding horizon) imposes a strong constraint on movement production: movements comply with a universal relationship between duration and jerkiness (Salmond et al., 2017; Shmuelof et al., 2012). Avoiding the production of unrealistic movements (e.g. smooth long-duration movements) probably contributes to the performance of the theory. The third principle is more enigmatic. It was adopted to explain velocity fluctuations during slow movements (Guigon et al., 2019) and seems effective to specify the kinematics and duration of a large class of movements. In fact, updating of goals at ~8 Hz is equally efficient to produce point-to-point movements of different durations and arbitrary drawing movements, and account for their properties (scaling law, speed/accuracy tradeoffs, power laws). The reason for this is unclear but it is satisfying to note that no additional hypothesis was needed to extend the model beyond the scope of slow movements.

## Relation to previous models

The current theory is based on two concepts (intermittency, receding horizon) which are used in the modeling framework developed by Bye and Neilson (2008, 2010). In their formalism, intermittency corresponds to successive 100-ms refractory periods. During a refractory period, processing of incoming information is delayed to the next period and a trajectory between predicted initial state and predicted target state is planned over a variable horizon. At the end of the refractory period, the planned trajectory is executed and a new refractory period starts in parallel. Accordingly, an action consists in a concatenated sequence of fixed duration submovements. Depending on the task, the variable horizon is either a receding horizon or a fixed horizon. For instance, a fixed horizon is used in tasks calling for temporal and spatial accuracy. On the one hand, our theory agrees with the view of Bye and Neilson that intermittent processes at a 100-ms time scale (close to $T_G$) are fundamental in motor control. For Bye and Neilson, the rationale for intermittency is to account for the well-documented psychological refractory period (Pashler, 1994). We could endorse this view since we have no independent reason for intermittency except that it helps accounting for the characteristics of kinematic fluctuations (Guigon et al., 2019). On the other hand, there is a clear difference between the variable horizon of Bye and Neilson which can be either a fixed or a receding horizon, is task-dependent, and varies as a parameter, and our receding horizon which is task-independent and is not a varying parameter. Bye and Neilson (2008) have shown that logarithmic and linear speed-accuracy tradeoffs rely on receding and fixed horizon control, respectively. In contrast, all speed-accuracy tradeoffs and their associated kinematics emerge from receding horizon control in our modeling framework.

## Limitations of the theory

A limitation of the theory is the versatility in task representation (series of goals, nature of boundary conditions) afforded by the third computational principle. A series of goals could be chosen by hand to obtain any desired motor behavior. Playing with all available states (and not solely position and velocity) would still increase the size of the space of representations. Thus, although the model has no free parameters, an open question is whether extended variations in implementation details and in the nature of noise are necessary to guarantee its explanatory power. A summary of all the simulations is given in Table 2. Several patterns appear. First, the full-state constraint applies to discrete point-to-point movements (gray lines). For these movements, partial constraints on the position or velocity of via-points lead to poorer results. The value method plays no specific role except for discrete Fitts' law. We could remove this notion and consider only the zero-value method. Second, continuous and rhythmic movements (white lines) involve the partial-position constraint and poorer results were obtained with other types of constraint. Third, the goal setting policy is a relevant dimension for discrete movements. Fourth, signal-dependent motor noise is central to movement variability although temporal noise is also required to explain temporal variability.

Another limitation is related to the (im)possibility to falsify the theory. We have presented all what the theory can explain, but what about aspects that cannot be explained? We have not found any property that we have tried to reproduce and for which we have obtained a wrong prediction. Yet as neither a confirmation bias nor bad faith can be definitely excluded, we need to be careful on this issue.

The scope of the theory is not known precisely. Although the theory was developed exclusively in relation to movements of the upper limbs, there is no reason why it would not apply to other bodily movements (leg, trunk, head, face, ...). Yet it is unclear how a task like locomotion can be cast in the proposed framework. It is also unclear whether it could apply to eye movements. Saccades share a number of properties with limb movements (kinematics, scaling laws, Fitts' law) and are successfully modeled in the framework of optimal control (Harris & Wolpert, 1998, 2006; Saeb, Weber, & Triesch, 2011; van Beers, 2008). However, there is a striking absence of isochrony in saccade production[13] which appears incompatible with the principles of the theory.

By construction, the theory leaves no room for a role of stiffness in the production of movements. As in many other computational models, the role of stiffness is thought to be restricted to a contribution to the compensation of external perturbations. Yet this view does not acknowledge the complex and refined contribution of stiffness to motor control (Burdet, Osu, Franklin, Milner, & Kawato, 2001). Recent models have attempted to give a more precise account of the role of stiffness in the framework of optimal control models (Berret & Jean, 2020; Mitrovic, Klanke, Osu, Kawato, & Vijayakumar, 2010).

## Debates on motor control

The debate about motor control has a large spectrum, but has in fact been centered on two Shakespearian dilemma: To control or not to control? To model or not to model? At one end of the spectrum, an ensemble of theoretical constructs (named below DA, *dynamical approach*) based on the ecological approach of Gibson (1966, 1986), physical theories of self-organized pattern formation (Haken, 1983) and dynamical system theory (Strogatz, 1994) claim that neither a centralized control nor a centralized representation are necessary for motor coordination, and behavioral motor patterns emerge from abstract, phenomenological nonlinear dynamics corresponding to interactions between the brain, the body and the environment (Kelso, 1995; Warren, 2006). At the other end, the *computational approach* views the nervous system (or part of it) as a computing device that builds models of the body and the environment, and controls changes in body states to achieve goals in the environment (Kawato, 1999; Todorov & Jordan, 2002). In between, the Passive Motion

---

[13] Yet, other eye movements such as eyelid movements (Evinger, Shaw, Peck, Manning, & Baker, 1984) and saccades without gaze-evoked blinks (Powers, Basso, & Evinger, 2013) are isochronous.

Paradigm[14] (PMP; Mohan & Morasso, 2011) combines classical feedback control and forward modeling, the Equilibrium Point Theory (EPT; Feldman & Levin, 1995) is closely related to classical feedback control theory[15], but denies the existence of internal models, and the Active Inference theory (AIT) exploits the notion of internal (generative) model, but views motor control as an inference problem rather than a control problem (Friston, 2011). Available discussions have in general been restricted to these dilemma, and have failed to provide decisive, convincing arguments for or against a particular theory.

Interestingly, a line of reasoning based on the distinction between position and force control (Ostry & Feldman, 2003) lead to sharper and less consensual arguments. On the one hand, models of movement control based on the direct specification of forces fail to properly take into account the existence of posture-stabilizing mechanisms (von Holst & Mittelstaedt, 1950/1973), and incorrectly predict patterns of muscular coactivation for postural maintenance following a displacement (Ostry & Feldman, 2003). On the other hand, production of movement based on position control is intrinsically compatible with postural control as a direct consequence of a postural resetting mechanism (Feldman & Levin, 1995). In DA, EPT, PMP and AIT, the basic unified mechanism of movement production and postural control is determined by an autonomous dynamical system, and is based on a local fixed-point dynamics[16], i.e. a restoring force[17] is exerted which is a function of the distance to a (static or moving) equilibrium point with a constant or state-dependent (possibly nonlinear) gain (stiffness). This description fits with the notion of position control. According to these theories, postural maintenance should correspond to such fixed-point dynamics. Yet, there is strong evidence that posture cannot be uniquely described by this kind of process. For whole body posture, ankle stiffness is in general too low to guarantee passive stabilization, which calls for the existence of an active, predictive postural control mechanism (Amiri & Kearney, 2020; Bottaro, Casadio, Morasso, & Sanguineti, 2005; Casadio, Morasso, & Sanguineti, 2005; Loram, Kelly, & Lakie, 2001; Loram & Lakie, 2002; Morasso & Schieppati, 1999; Morasso & Sanguineti, 2002). Studies on the manual stabilization of unstable objects revealed two strategies: a high-stiffness strategy and a low-stiffness, intermittent, predictive strategy (Lakie, Caplan, & Loram, 2003; Saha & Morasso, 2012). These observations indicate that theories based on position control would need to be amended to account for the need of predictive control, which in fact would violate their very principles. The computational approach is less sensitive to this debate between position and force control because it is versatile and can be cast in different ways, e.g. as predictive control of a positional variable (Guigon, 2010).

---

[14] Whether the PMP should be considered as a "computational" method or not is discussed in Mohan and Morasso (2011).
[15] The control theoretical framework is not intrinsically incompatible with "non-computational" approaches, e.g. some dynamical models can be cast in terms of the control theory (Sternad, 2000, pp 415-416).
[16] In the dynamical approach, the dynamics can also be governed by a limit-cycle attractor.
[17] In most cases, a second-order dynamics is used which allows to use the term "force". Yet this term is not quite appropriate for an abstract dynamics.

The ultimate goal of a theorist is, on the one hand, to design an overarching theory in his field of research and, one the other hand, to elaborate an argumentative discourse that supports his theory and can dismiss alternative, concurrent proposals. Although the two tasks are important on their own, the latter should not be used shamelessly to conceal difficulties in the former. At its inception (Kugler, Kelso, & Turvey, 1980), the dynamical approach was conceived as a formal alternative to the cognitive/information-processing theory of action, but with no proper content. With the advent of the finger "wipers" task (Kelso, 1984; Kelso, Holt, Rubin, & Kugler, 1981), ensuing modeling (Haken, Kelso, & Bunz, 1985) and later developments (review in Beek, Peper, & Stegeman, 1995; Kelso, 1995), the dynamical approach became a well-established theory of action encompassing behavioral and neural levels (Jirsa, Friedrich, Haken, & Kelso, 1994; Schöner, 1990, 1994). Yet its explanatory power remains largely limited to variations around the original task (Sternad, 1998, 2000).

Recent works (Huys et al., 2014; Warren, 2006) involved long arguments and long methodological developments, but produced little insights outside the historical scope of the theory. In the meantime, several critical issues remain unsolved. The first issue is related to the possibility that common principles apply to rhythmic and discrete actions (Beek et al., 1995; Daffertshofer, van Veen, Ton, & Huys, 2014). The second issue concerns the lack of a general method to guide the design of a dynamical system for reproducing a given behavior (Ijspeert, Nakanishi, Hoffmann, Pastor, & Schaal, 2013). A possible solution would be to use computational tools (optimization, machine learning techniques) to discover or learn appropriate dynamics (Ijspeert et al., 2013; Schaal, Mohajerian, & Ijspeert, 2007), but it is not in the spirit of the dynamical approach. The third issue is related to the very nature of emergent trajectories defined in general in an abstract space (e.g. relative phase in Haken et al., 1985; but for a definition in task or body space, see Haken et al. 1985; Saltzman & Kelso, 1987; Schaal et al., 2007) and how they can be translated into actual displacements of a necessarily mechanical object (e.g. a finger). The trajectories should thus be considered as an input to a trajectory-following system. At this stage, the difficulty is not solely related to the lack of flexibility of trajectory-based control (Schaal et al., 2007), but more deeply to the fact that remaining redundancies below the level of trajectory definition (e.g. in muscle or neural space) are neither considered nor exploited (Todorov & Jordan, 2002). The fourth issue is the intrinsic temporal and spatial invariance of the solutions of autonomous dynamical systems which can be interesting in robotics (Schaal et al., 2007), but is not consistent with human motor behavior (Gentner, 1987; Smith, Goffman, Zelaznik, Ying, & McGillem, 1995). The last two issues also apply to the Equilibrium-Point Theory (Feldman & Levin, 1995) and the Active Inference theory (Friston, 2011; Friston et al., 2011) since they are cast in terms of a trajectory-following problem. The Passive Motion Paradigm is by construction more computational and versatile than DA, EPT and AIT, and in fact, by different aspects, not so far from optimal control models (Mohan & Morasso, 2011). Yet this theory would need to be further develop toward an overarching account of motor control.

## Conclusion

In keeping with the theory of Todorov and Jordan (2002), the present theory explains characteristics of motor control as by-products of computational principles (optimal feedback control, receding horizon, task representation by a series of goals). None of the principles is new (Guigon et al., 2019; Todorov & Jordan, 2002), but their assemblage in an overarching theory of motor control is new. These principles are sufficient to provide a detailed account of a large set of motor behaviors (discrete, continuous, rhythmic and timing actions) and properties (scaling laws, power laws, speed-accuracy tradeoffs). The theory significantly extends our understanding of action production in the framework of control theory.

# Tables

| Time | Explanation |
|---|---|
| $\Delta$ | duration of a simulation |
| $t$ | current time during the simulation |
| $\delta$ | timestep of the simulation — for numerical integration |
| $T_H$ | receding horizon — in the model the value of $T_H$ is constant (0.28 s) |
| $T_0$ | fixed horizon — used only to contrast with the receding horizon |
| $T(t)$ | time interval to reach a goal at time $t$ (Equation 1) — the value of $T(t)$ can be: $T_0$ (fixed horizon control); $\infty$ (infinite horizon control); or $t + T_H$ (receding horizon control) — only the receding horizon control is used |
| $T_G$ | period at which goals are updated in the absence of temporal constraints — in the model the value of $T_G$ is constant (0.13 s) |
| $T_{step}$ | period at which goals are updated in general (Figure 1C) — the value of $T_{step}$ is: $T_G$ in the absence of temporal constraints; any value in the presence of temporal constraints (e.g. period of a metronome) |
| $t_{k\ (k=1,\ldots N)}$ | times at which goals are updated (Equation 6) — $t_{k+1} - t_k = T_{step}$ |

**Table 1**. Summary on time notations.

| exp | constr | var | VP | TA | VM | GS | noise | Fig | comment |
|---|---|---|---|---|---|---|---|---|---|
| **fastest** | full-state | $p$ | — | (11) | — | — | — | 2 | — |
| **slow** | full-state | $pv$ | (12/14) | — | Z/C | A | — | 3 | Does not work with relative goal setting. Works with partial pv constraints. |
| **interm.** | full-state | $p$ | (11) | (11) | Z/C | A/R | — | 4 | — |
| **drawing** | partial position | $p$ | (9) | — | — | A | $SDN_m$ | 5,6 | Poorer fit with partial pv and full-state constraints, and with relative goal setting. |
| **writing** | partial position | $p$ | (9) | — | — | A | $SDN_m$ | 7 | Poorer fit with partial pv and full-state constraints, and with relative goal setting. |
| **scaling laws** | full-state | $p$ | (11) | (11) | Z/C | A | — | 8,9 | Poorer scaling with partial p and pv constraints. |
| | | | | | | R | — | 8,9 | Loss of invariance of time to peak velocity for partial p and pv constraints. |
| **rhythmic** | partial position | $p$ | — | (9) | — | — | $SDN_m$ | 10 | Qualitatively similar results with partial pv and full-state constraints. |
| **discrete Fitts** | full-state | $p$ | (11) | (11) | Z | A | $SDN_m$ | 11 | Loss of Fitts' law with current-value method. Fundamental role of $SDN_m$. |
| | | | | | | R | $SDN_m$ | 11 | Loss of Fitts' law with current-value |

| | | | | | | | | | method. |
|---|---|---|---|---|---|---|---|---|---|
| **rhythmic Fitts** | partial position | $p$ | — | (9) | — | — | $SDN_m$ | 12,13 | Loss of Fitts' law with partial pv and full-state constraints. |
| **timing** | full-state | $p$ | (11) | (11) | Z/C | R | $T_G$ | 14,15 | Importance of temporal noise to account for temporal variability. |

Table 2. Comparison between the simulations. **constr**: boundary constraint; **var**: variable; $p$: position; $v$: velocity; **VP**: via-point; **TA**: target; **VM**: value method; Z: zero-value; C: current-value; **GS**: goal setting policy; A: absolute; R: relative. Number in parentheses refer to equations. Gray lines indicate discrete movements.

# Appendix A — Parametric study of power laws

A parametric analysis of the power laws obtained in drawing (Figure 6) has been performed. It involved $T_H$, $T_G$, and the nature of boundary constraints. It shows that $T_H = 0.28$ s, $T_G = 0.13$ s, and the partial-position constraint give the best fit to observed data (Figure A1), which provides an independent cross-check for $T_H$ and $T_G$.

*Figure A1*. Parametric study of drawing power laws. **A**. Influence of $T_H$ for $T_G = 0.13$ s and partial-position constraint: (*black*) $T_H = 0.28$ s; (*red*) $T_H = 0.18$ s; (*green*) $T_H = 0.38$ s; (*blue*) $T_H = 0.48$ s. **B**. Influence of $T_G$ for $T_H = 0.28$ s and partial-position constraint: (*black*) $T_G = 0.13$ s; (*red*) $T_G = 0.18$ s; (*green*) $T_G = 0.23$ s. **C**. Influence of boundary constraints for $T_H = 0.28$ s and $T_G = 0.13$ s: (*black*) partial-position constraint; (*red*) partial-position/velocity constraint with current-value method; (*green*) partial-position/velocity constraint with zero-value method; (*blue*) full-state constraint with current-value method; (*purple*) full-state constraint with zero-value method.

# Appendix B — Choice of via-points

*Figure B1*. Via-points. **A**. Position of the via-points for the point-to-point movements of Figure 8B. Absolute goal setting policy. **B**. Position of the via-points for the point-to-point movements of Figure 8C. Relative goal setting policy. (*diamond*) start point; (*plain square*) via-point; (*circle*) target.

# Appendix C — Phase transition in rhythmic movements

A cardinal observation in the study of rhythmic movements is the frequency-dependent phase transition in bimanual coordination (Kelso, 1984), i.e. synchronized, asymmetric (parallel) finger displacements are suddenly replaced by symmetric displacements as synchronization frequency increases. A popular explanation of this phenomenon is provided by the Haken-Kelso-Bunz model in terms of phase transition in a parameterized nonlinear dynamical system (Haken et al., 1985). The predictive power of this model is undeniable and striking (Kelso, 1995), yet its special-purpose nature is a matter of concern (Ijspeert et al., 2013). Todorov and Jordan (2002) proposed a partial account of phase transition in the framework of optimal feedback control. They observed that in a one-dimensional sinusoidal tracking task with a redundant, 2-dof (telescopic) arm, the phase between the first segment of the arm and the endpoint changed abruptly with frequency (at ~2.8 Hz)[18]. The present theory could endorse this view to extend its explanatory power. Yet an alternative approach can be developed based on the notion of task representation. In the study of drawing movements, we have shown that the fastest two-dimensional drawing defined by internally updated via-points has 3 via-point and lasts ~0.38 s, i.e. ~2.6 Hz for repetitive drawing (Figure 5E, inset). In this framework, the only mean to increase movement frequency is to make a one-dimensional drawing (2 via-points). If we consider that asymmetric and symmetric patterns abstractly correspond to 2D and 1D representations, respectively, phase transition is explained by the necessary change of dimension to increase movement frequency. A phase transition between drawing the figure "eight" and drawing the figure "zero" has been described by Buchanan, Kelso, & Fuchs (1996) with a critical frequency between 1.2 and 1.4 Hz. In the model, the figure "eight" can be drawn with different numbers of via-points (here from 8 to 4;

---

[18] Experimentally measured transition frequencies vary considerably between participants (Kelso, 1984). In Scholz, Kelso, & Schöner (1987), the range is 2-2.6 Hz.

Figure C1A,B,D,E). The minimum number of via-points is 4 (which can also be used to draw the figure "zero"; Figure C1F) corresponding to a frequency of 1.89 Hz. The discrepancy between the experimental and theoretical transition frequencies may be related to what is considered as a proper "eight" representation. For instance, a transition could occur around 1.28 Hz, between 8 and 6 points, if the 6 points are interpreted as a "zero" rather than as a "eight" (Figure C1A,C). Overall, this view of phase transition is consistent with the perceptual rather than motoric nature of bimanual coordination (Mechsner, Kerzel, Knoblich, & Prinz, 2001).

> *Figure C1*. Via-point representations for drawing the figure "eight" and the figure "zero" (*left column*) and actual drawing (right column). The starting point is indicated by a open square, 50 turns. The actual number of via-points may be different from the observed number of via-points since some via-points can be used several times. Simulations with partial-position constraint, absolute goal setting policy, in the presence of $SDN_m$ ($\sigma^\varepsilon = 1$).

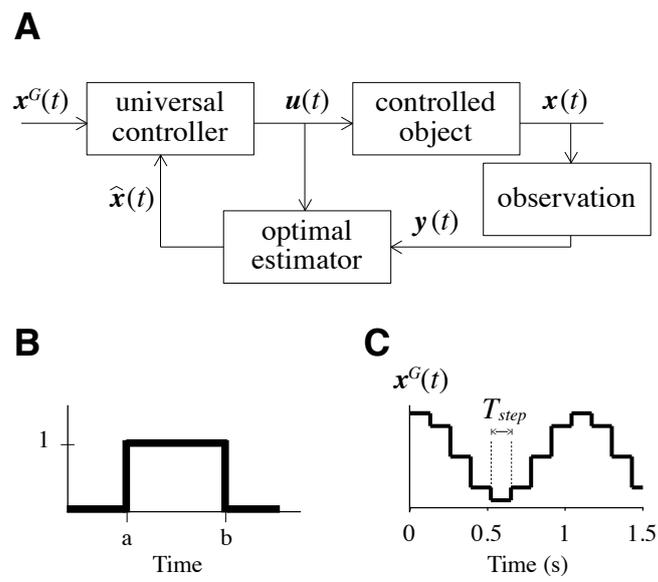

Figure 1

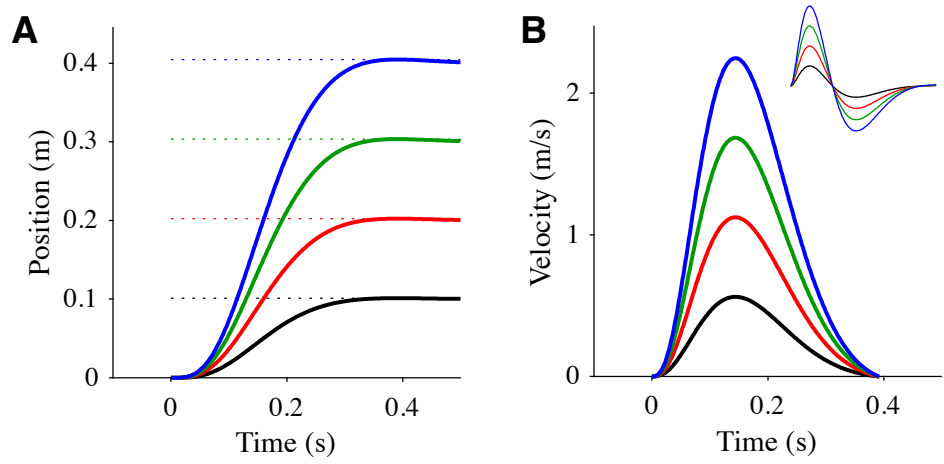

Figure 2

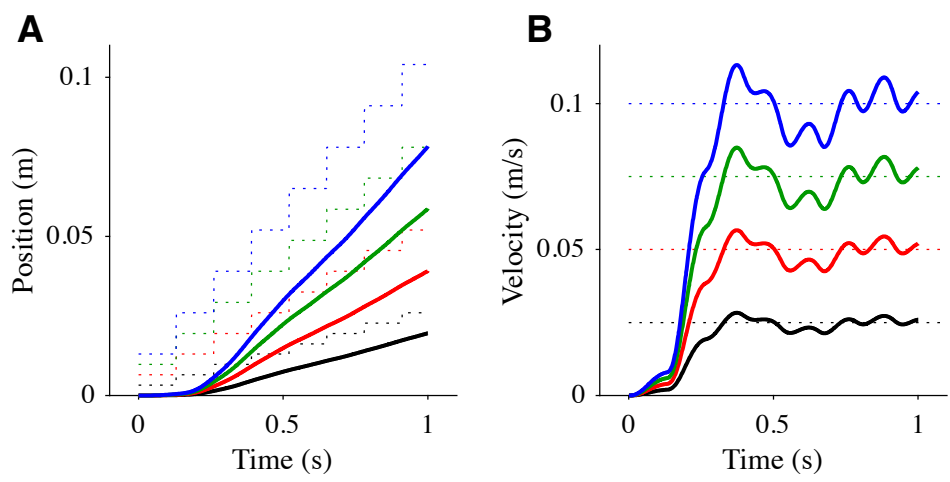

Figure 3

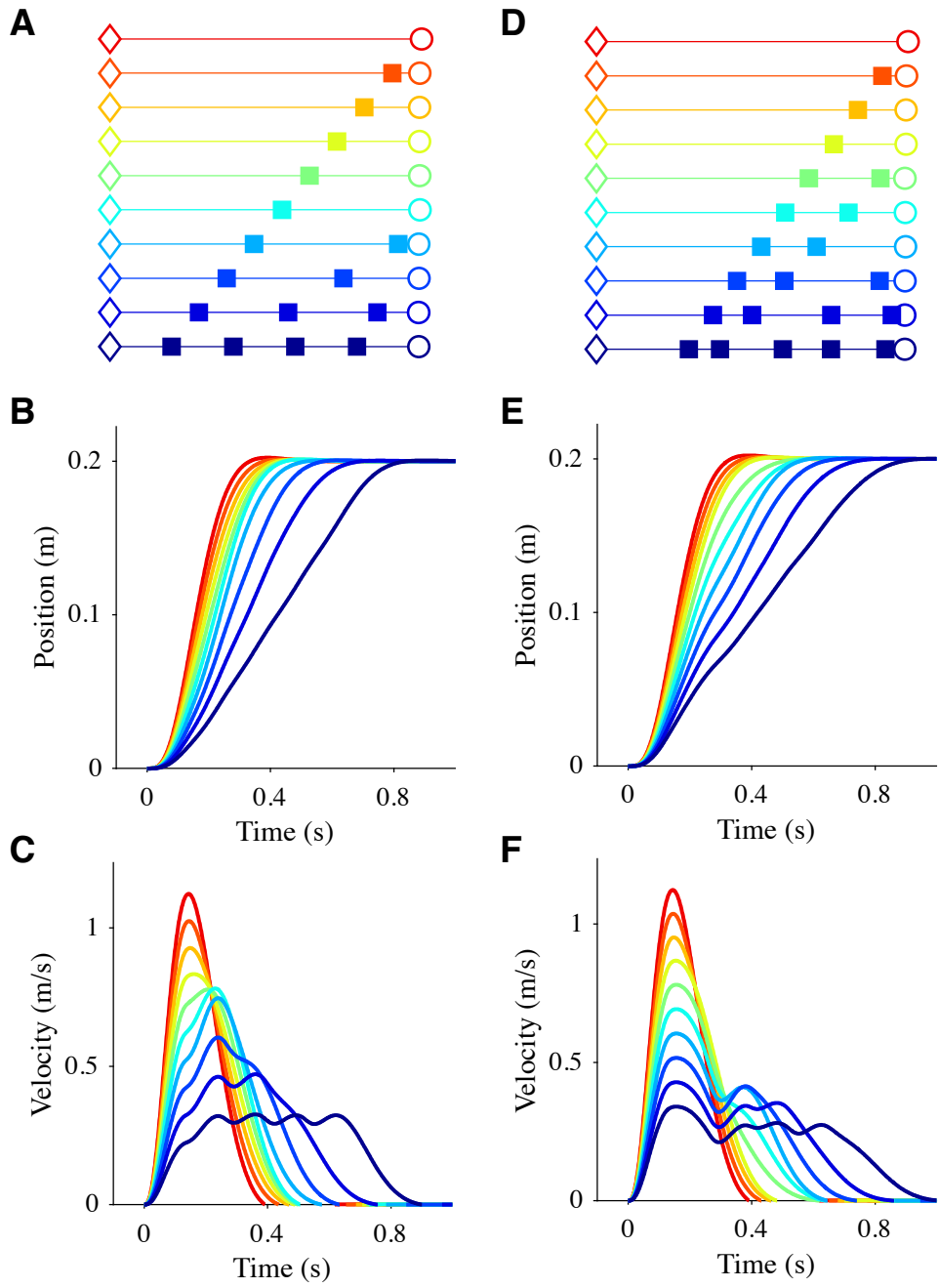

Figure 4

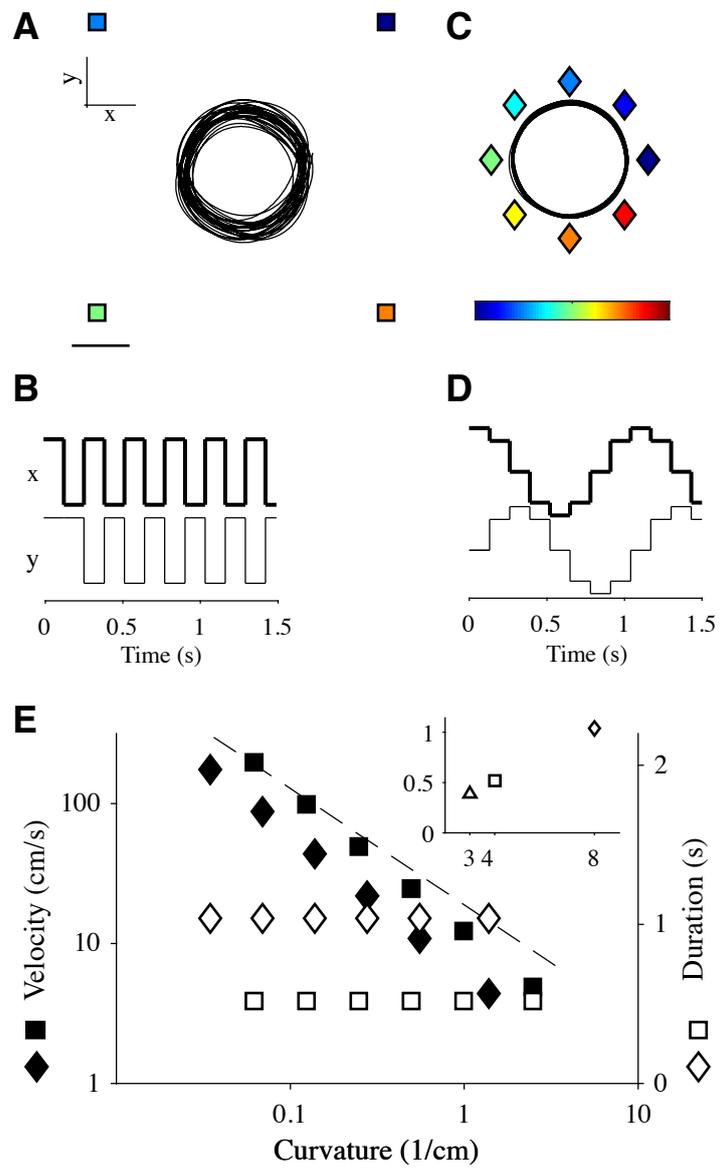

Figure 5

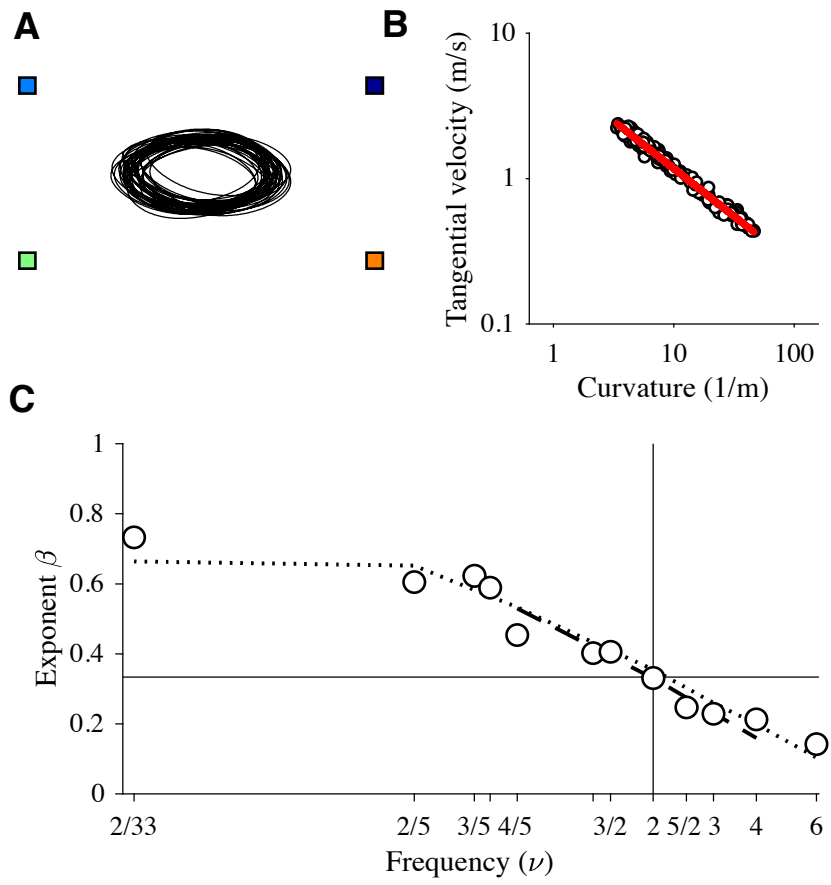

Figure 6

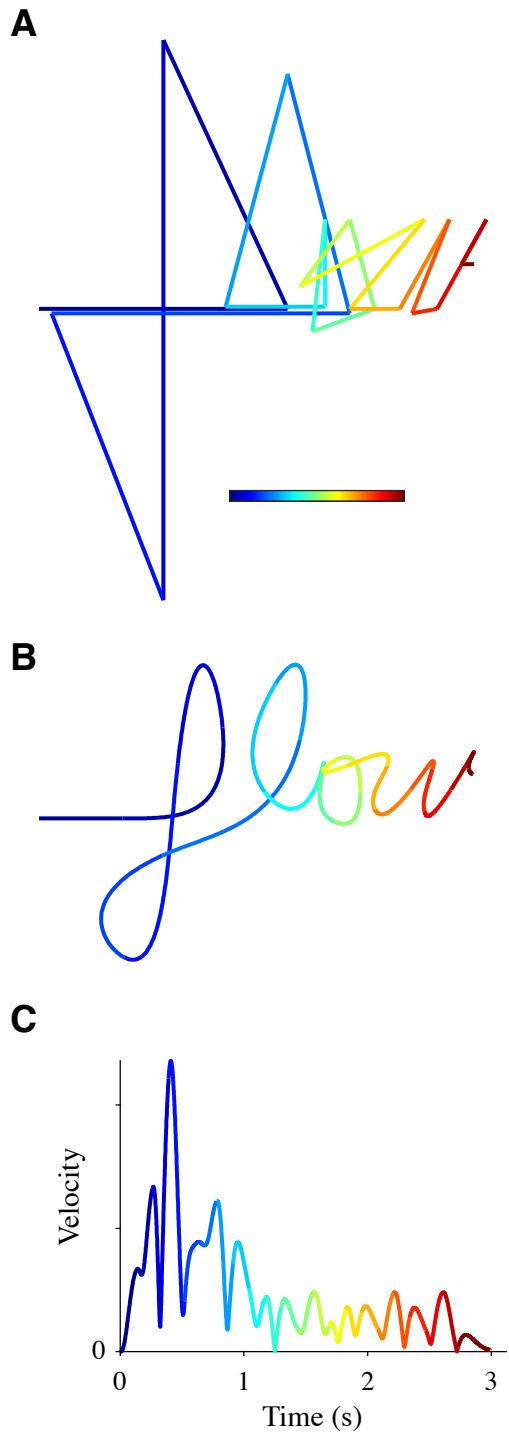

Figure 7

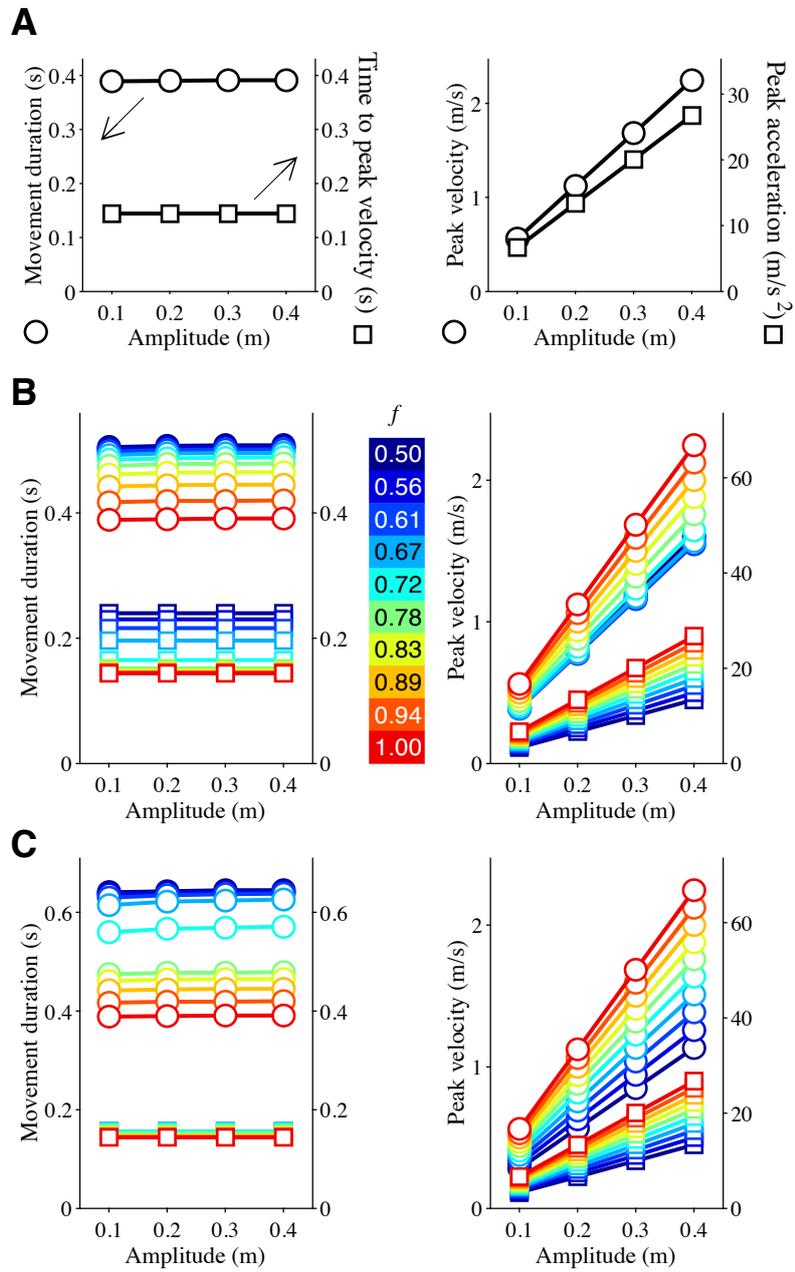

Figure 8

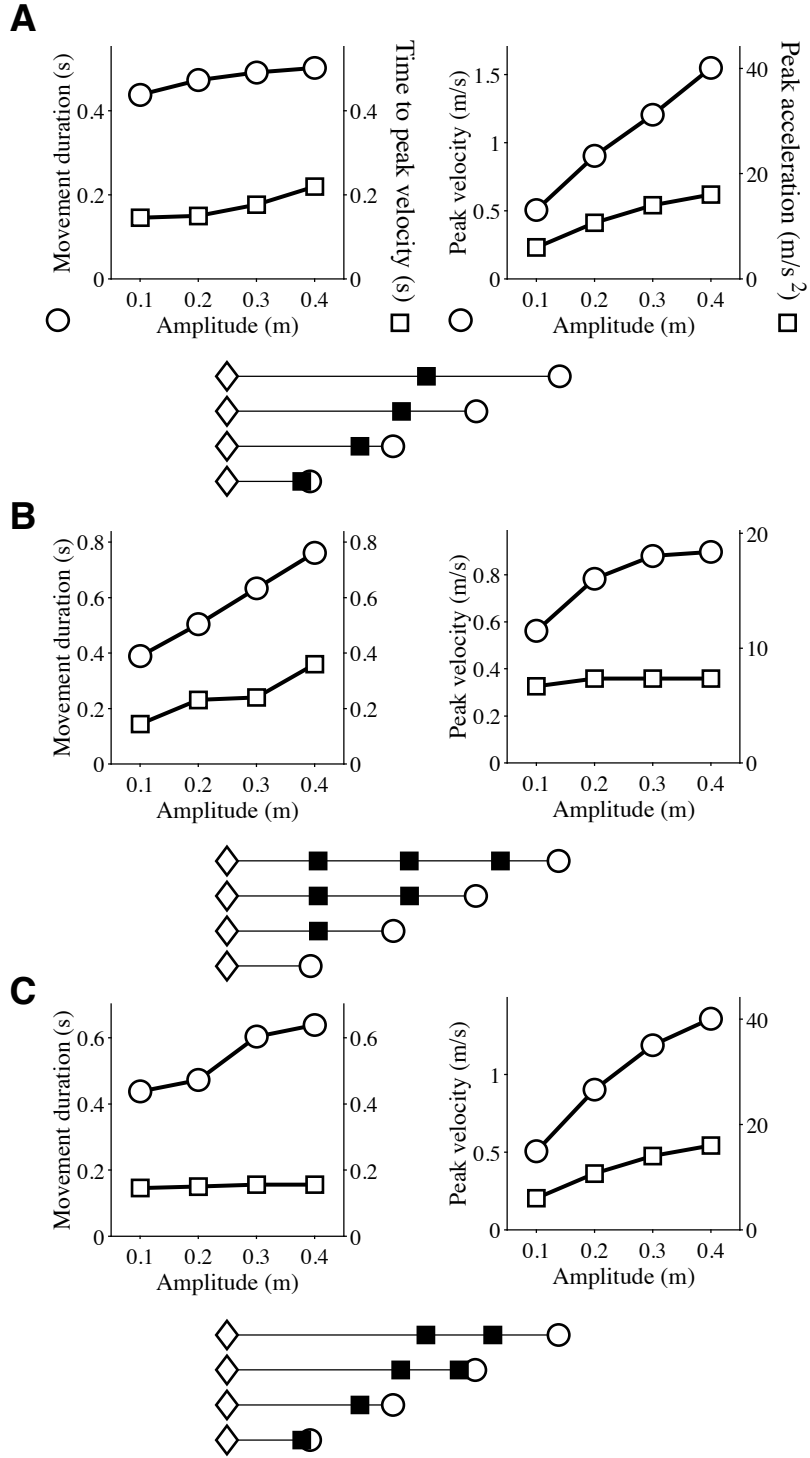

Figure 9

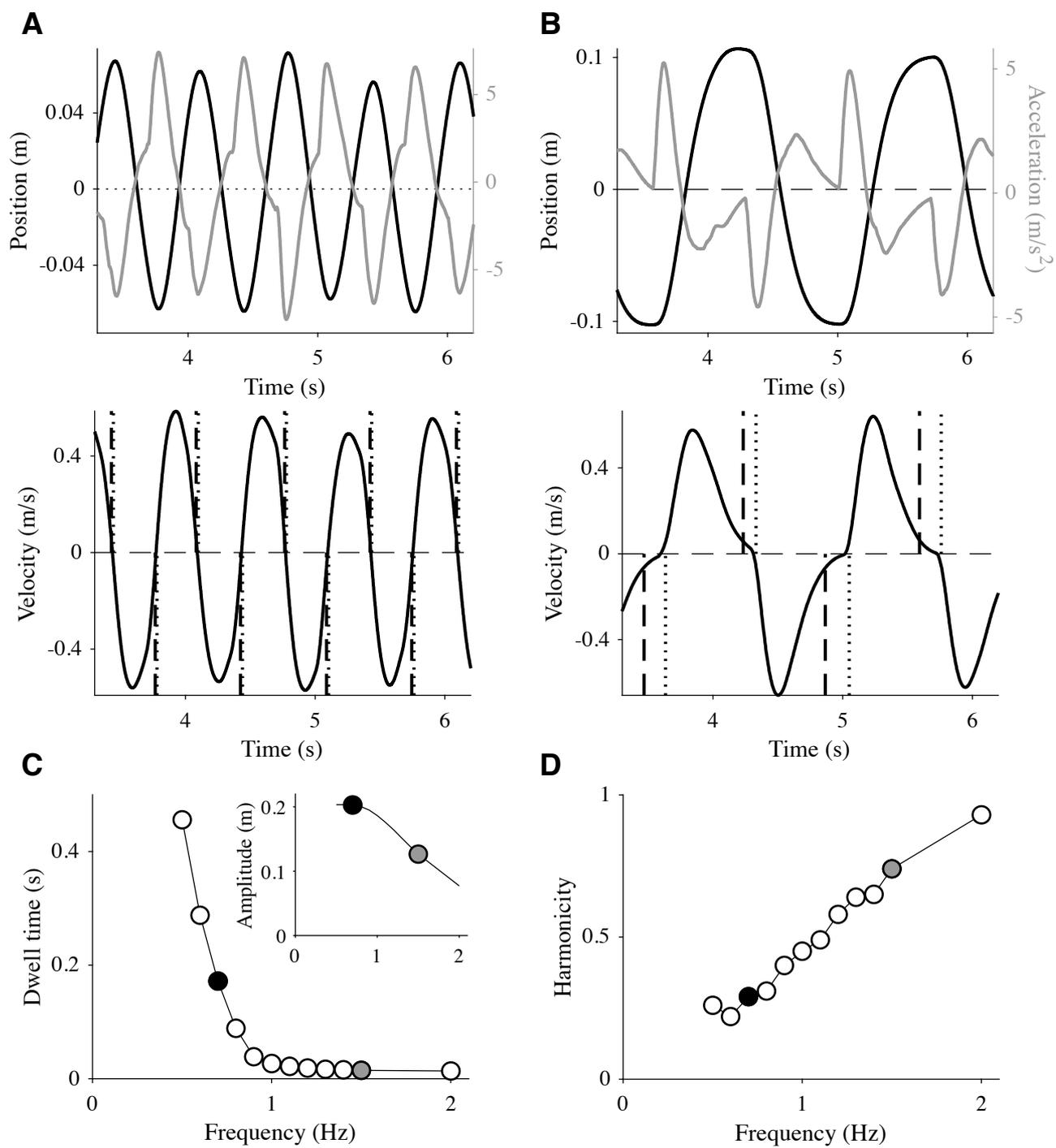

Figure 10

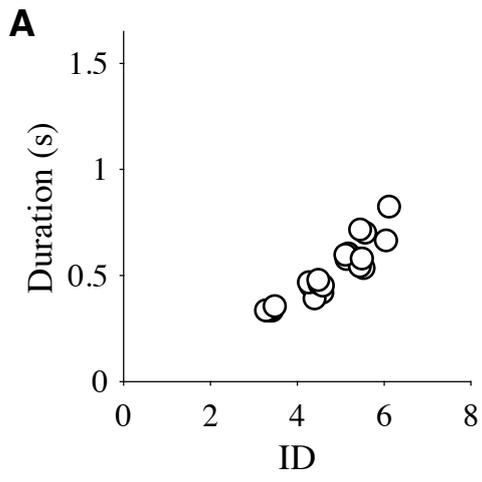
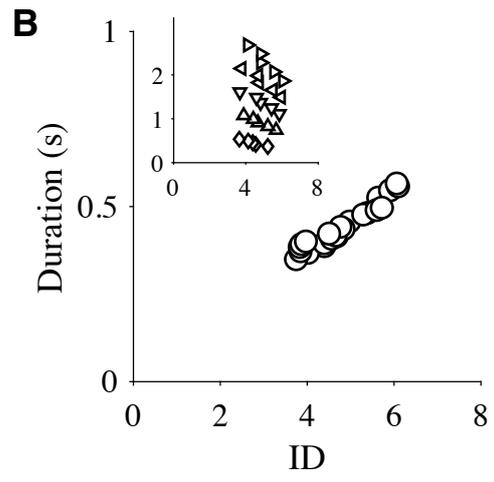
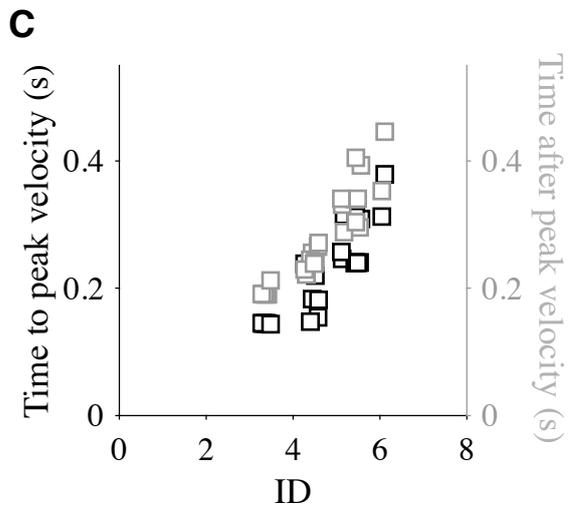
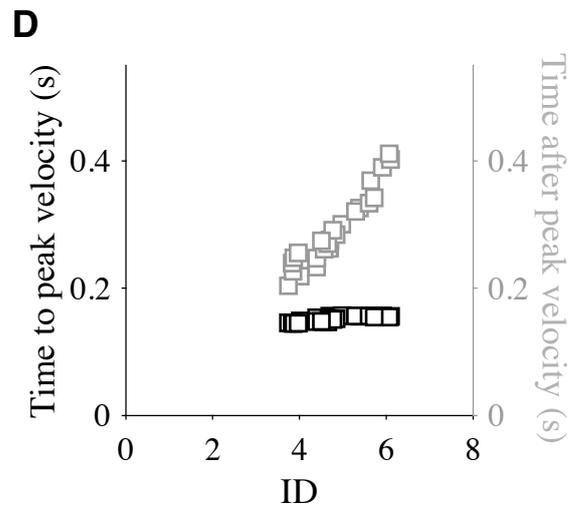

Figure 11

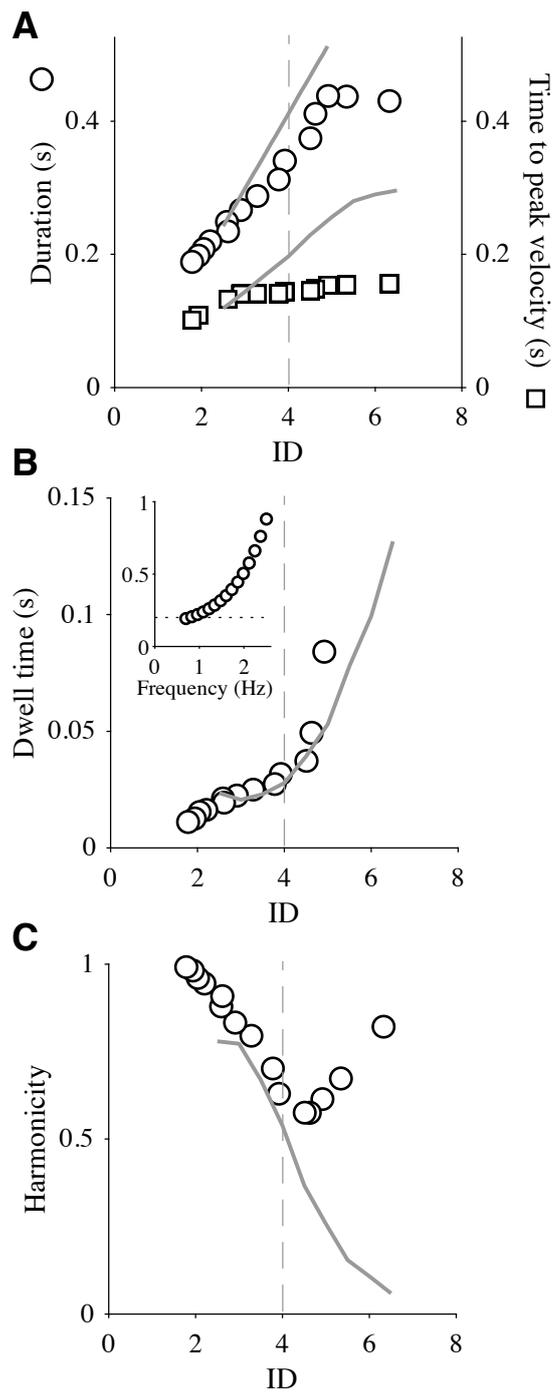

Figure 12

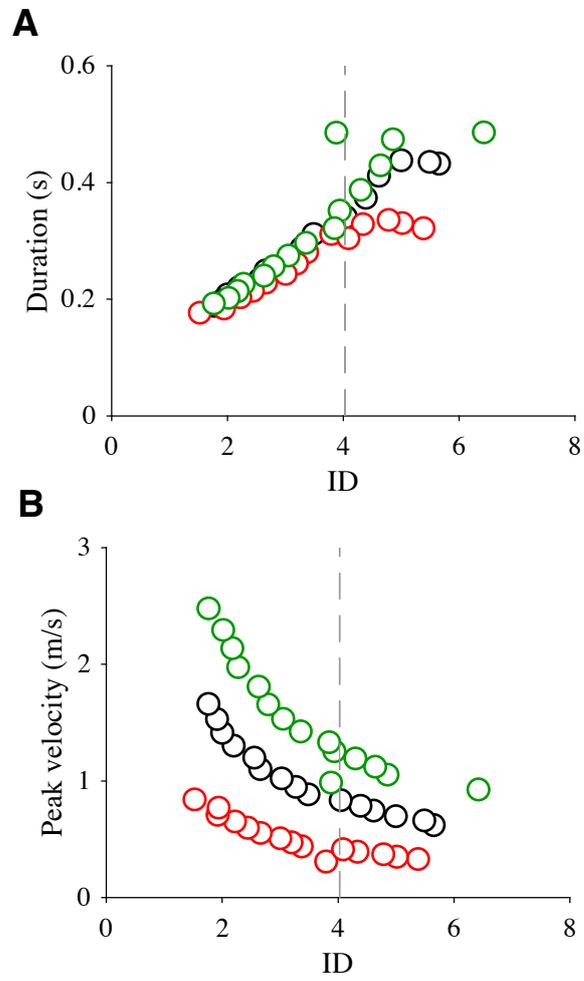

Figure 13

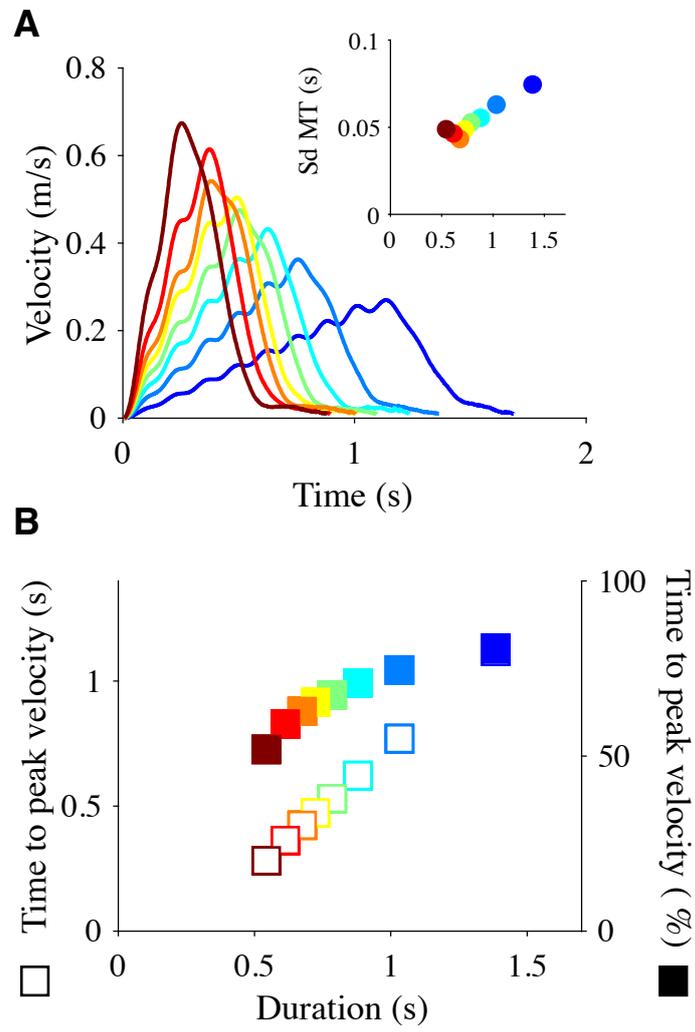

Figure 14

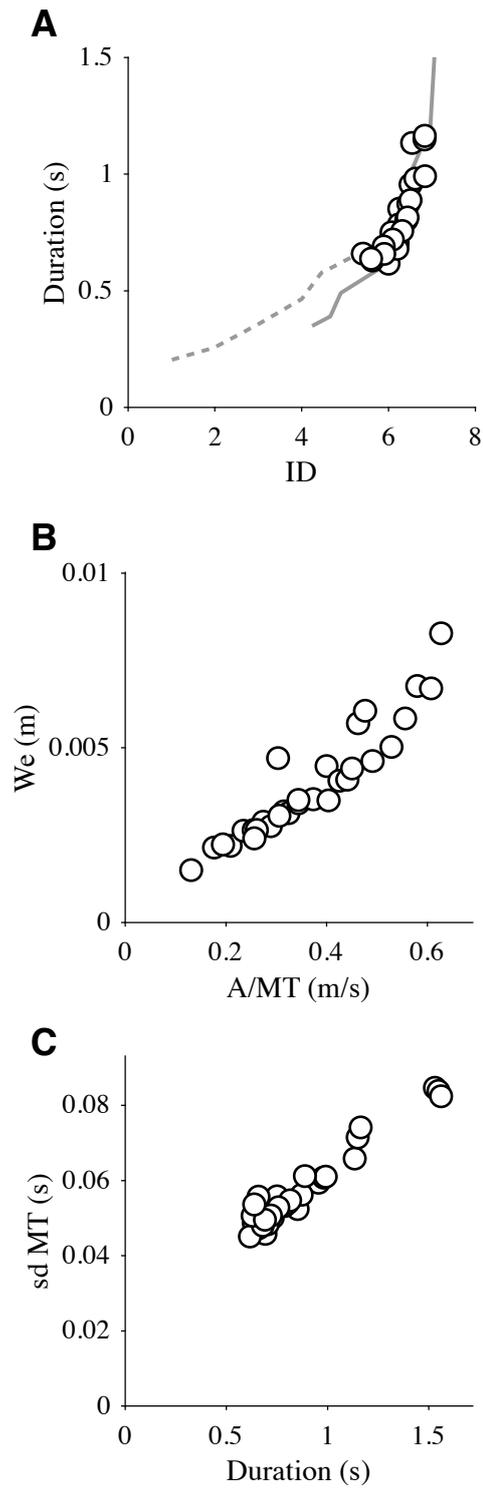

Figure 15

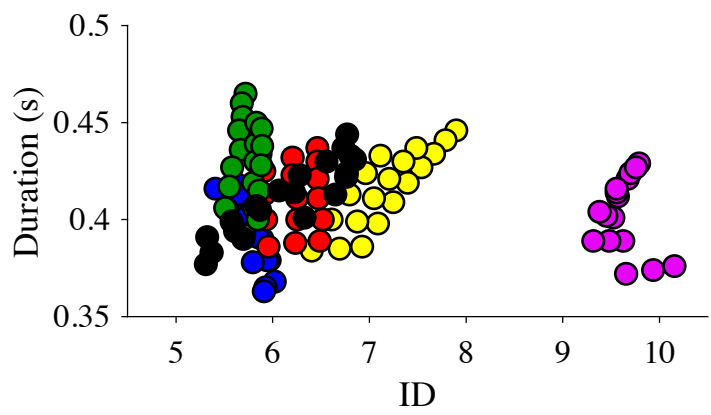

Figure 16

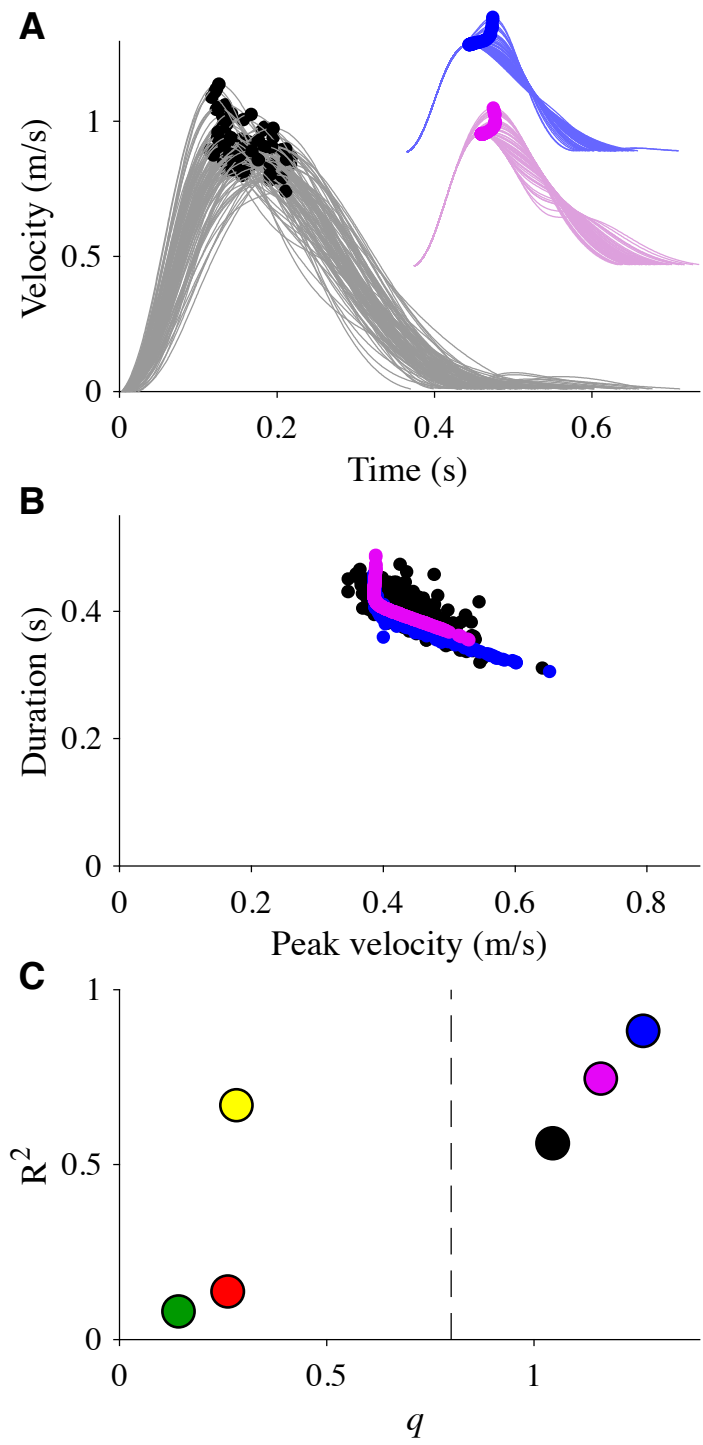

Figure 17

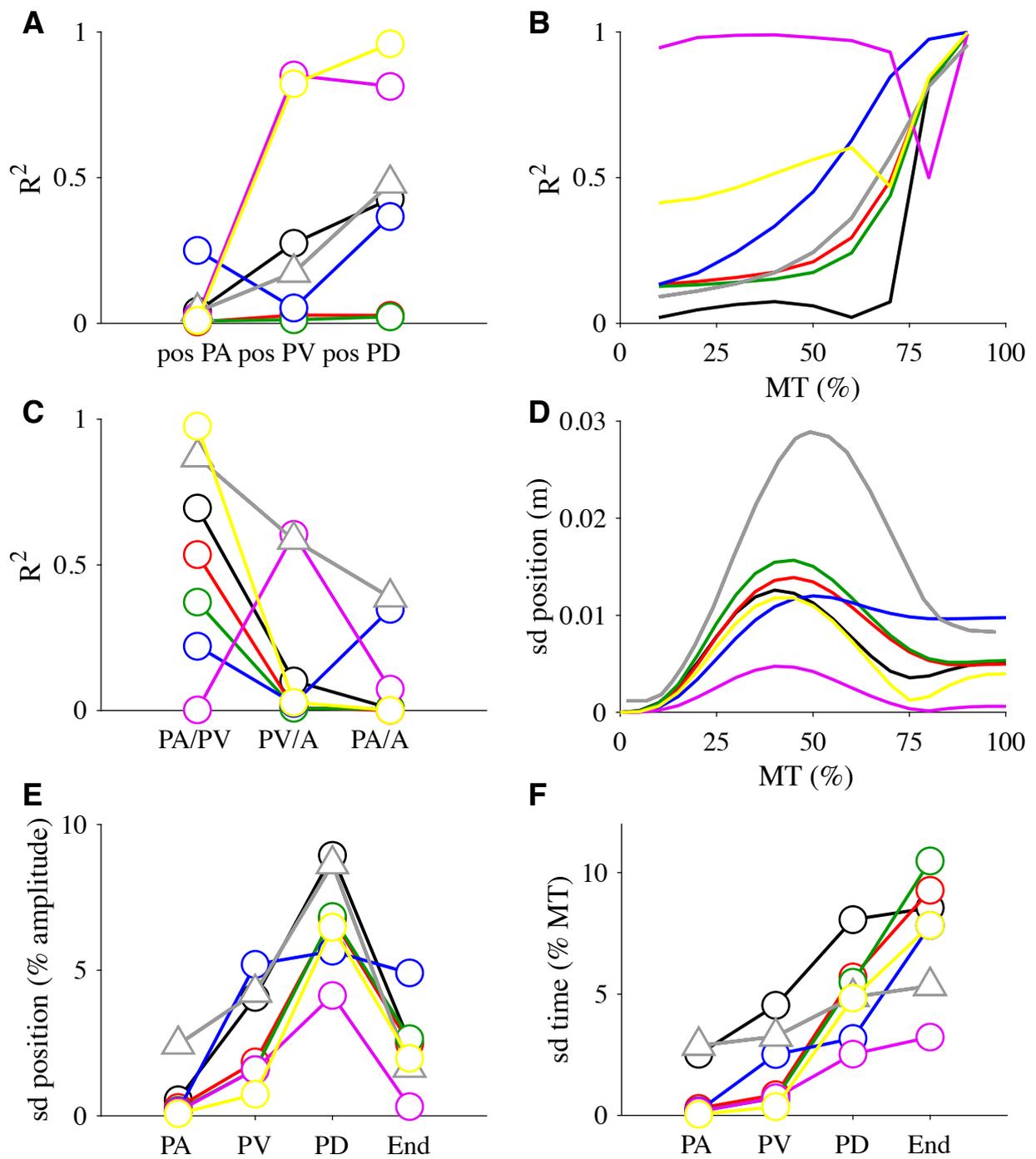

Figure 18

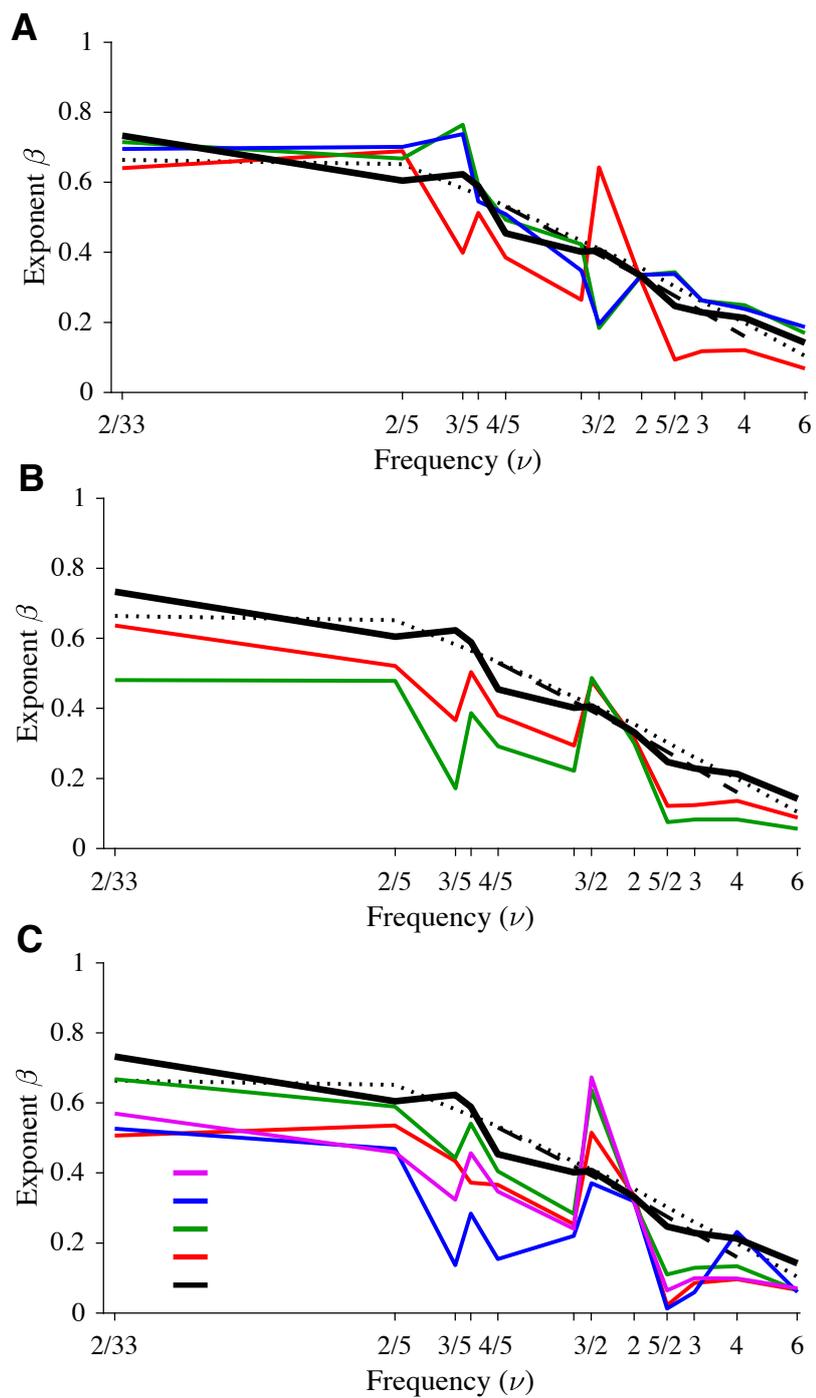

Figure A1

A 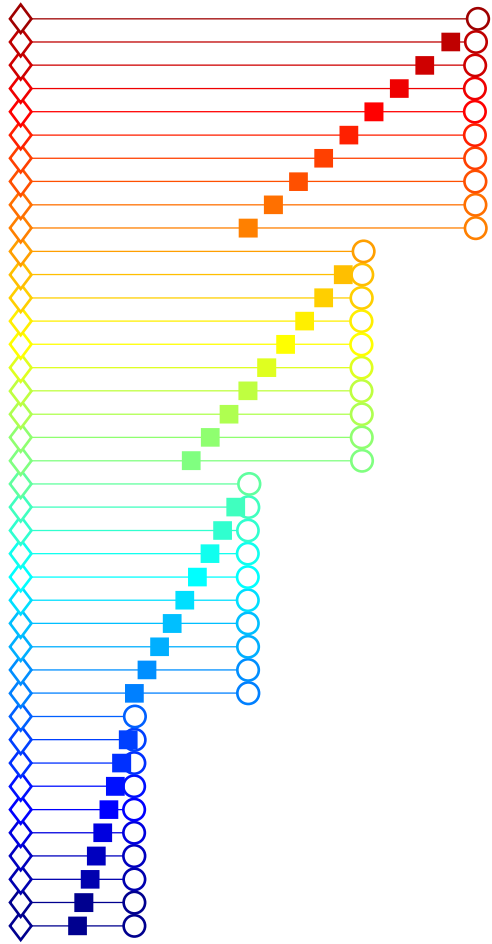 B 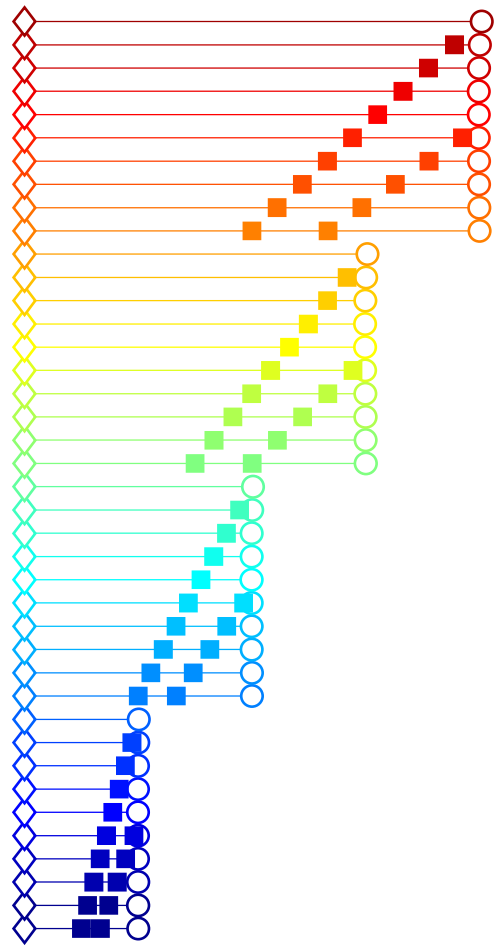

Figure B1

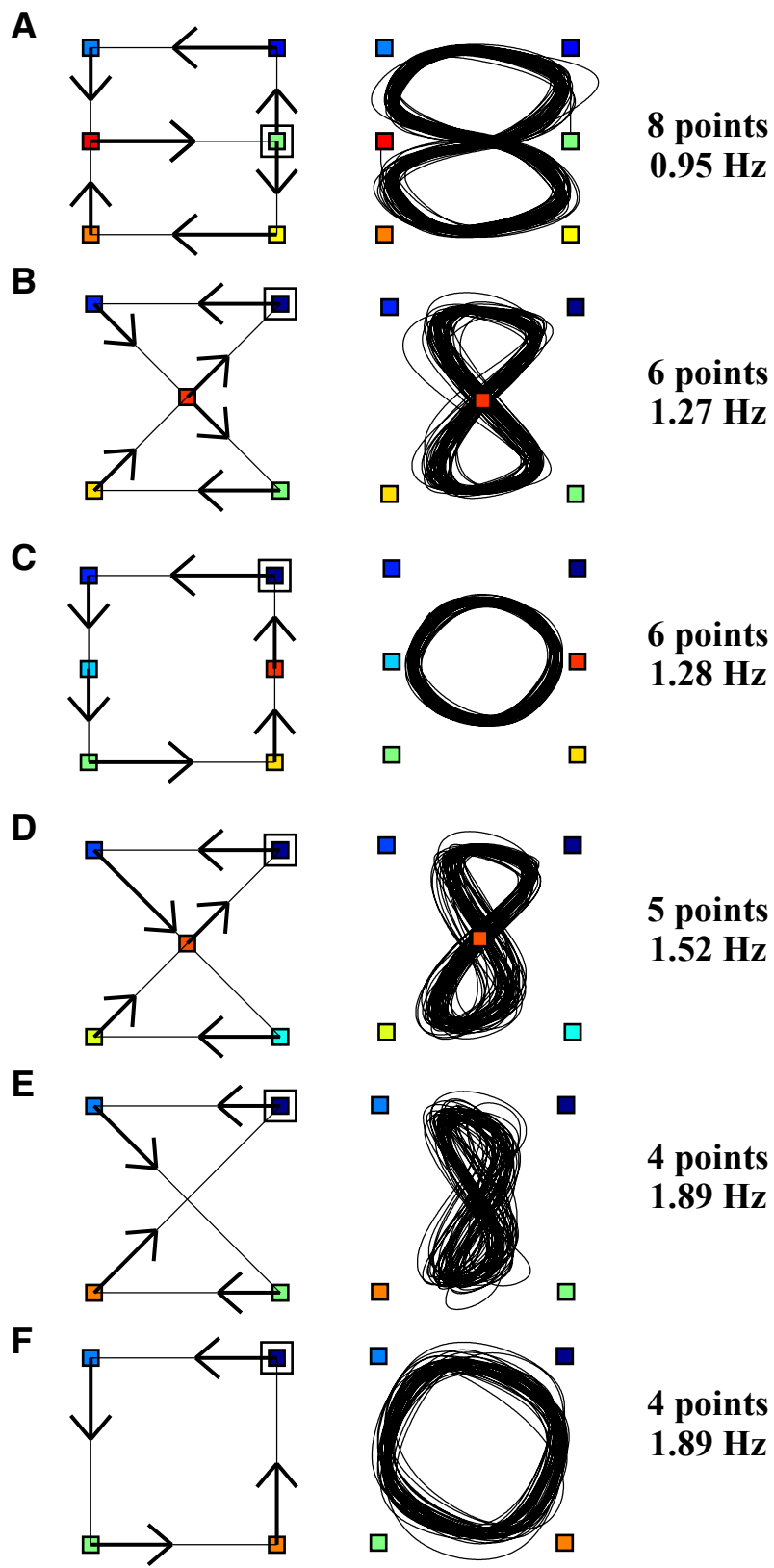

Figure C1

# Online Supplemental Material

# A computational theory for the production of limb movements

Emmanuel Guigon

March 12, 2020

Here, we present the mathematical background necessary to understand the proposed model.

## Optimal control with terminal constraints

We consider a dynamical system

$$\dot{\boldsymbol{x}}(t) = f\left[\boldsymbol{x}(t), \boldsymbol{u}(t)\right] \tag{S1}$$

where $\boldsymbol{x} \in \mathbb{R}^n$ is the state of the system and $\boldsymbol{u} \in \mathbb{R}^m$ a control vector. An optimal control problem for this system is to find the control vector $\boldsymbol{u}(t)$ for $t \in [t_0; t_f]$ to minimize a performance index

$$J = \int_{t_0}^{t_f} L\left[\boldsymbol{x}(t), \boldsymbol{u}(t)\right]\ dt \tag{S2}$$

with boundary conditions

$$\boldsymbol{x}(t_0) = \boldsymbol{x}_0 \qquad \psi\left[\boldsymbol{x}(t_f)\right] = 0. \tag{S3}$$

This problem is the generic formulation corresponding to Equations 1,2,3 of the article.

**Mayer formulation**

We first show that the optimal control problem defined by Eq. S1, Eq. S2 and Eq. S3 can be equivalently written

$$\dot{\tilde{\boldsymbol{x}}}(t) = \tilde{f}\left[\tilde{\boldsymbol{x}}(t), \boldsymbol{u}(t)\right] \tag{S4}$$

$$\tilde{J} = \phi\left[\tilde{\boldsymbol{x}}(t_f)\right] \tag{S5}$$

$$\tilde{\boldsymbol{x}}(t_0) = \tilde{\boldsymbol{x}}_0 \qquad \tilde{\psi}\left[\tilde{\boldsymbol{x}}(t_f)\right] = 0 \tag{S6}$$

which is called the Mayer formulation and which is simpler for numerical methods.

We consider the supplementary state variable $z$ defined by

$$\dot{z}(t) = L\left[\boldsymbol{x}(t), \boldsymbol{u}(t)\right]$$

and $z(t_0) = 0$. Thus $J = z(t_f)$. We define the new state variable

$$\tilde{\boldsymbol{x}}(t) = \begin{pmatrix} z(t) \\ \boldsymbol{x}(t) \end{pmatrix}.$$

We can reformulate the optimal control problem in the following way: find the control vector $\boldsymbol{u}(t)$ to minimize

$$\tilde{J} = \phi\left[\tilde{\boldsymbol{x}}(t_f)\right] = z(t_f) \tag{S7}$$

subject to

$$\dot{\tilde{\boldsymbol{x}}}(t) = \tilde{f}\left[\tilde{\boldsymbol{x}}(t), \boldsymbol{u}(t)\right] = \begin{pmatrix} L\left[\boldsymbol{x}(t), \boldsymbol{u}(t)\right] \\ f\left[\boldsymbol{x}(t), \boldsymbol{u}(t)\right] \end{pmatrix} \tag{S8}$$

and

$$\tilde{\boldsymbol{x}}(t_0) = \tilde{\boldsymbol{x}}_0 = \begin{pmatrix} 0 \\ \boldsymbol{x}_0 \end{pmatrix} \qquad \tilde{\psi}\left[\tilde{\boldsymbol{x}}(t_f)\right] = \begin{pmatrix} 0 \\ \psi\left[\boldsymbol{x}(t_f)\right] \end{pmatrix} = 0. \tag{S9}$$

Thus we can remove the integral term in the performance index. In the following we consider the problem defined by Eq. S4, Eq. S5 and Eq. S6. For simplicity, we remove the tilde sign.

**Solution**

We adjoin the constraints to the performance index with Lagrange multipliers $\boldsymbol{\nu}$ and $\boldsymbol{\lambda}(t)$

$$\bar{J} = \phi + \boldsymbol{\nu}^T \psi + \int_{t_0}^{t_f} \boldsymbol{\lambda}^T(t) \left\{ f\left[\boldsymbol{x}(t), \boldsymbol{u}(t)\right] - \dot{\boldsymbol{x}}(t) \right\} \, dt.$$

The Hamiltonian function is

$$\mathcal{H}[\boldsymbol{x}(t), \boldsymbol{u}(t), \boldsymbol{\lambda}(t)] = \mathcal{H}(t) = \boldsymbol{\lambda}^T(t) f[\boldsymbol{x}(t), \boldsymbol{u}(t)]. \tag{S10}$$

The generalized performance index can be written

$$\bar{J} = \Phi[\boldsymbol{x}(t_f)] - \boldsymbol{\lambda}^T(t_f)\boldsymbol{x}(t_f) + \boldsymbol{\lambda}^T(t_0)\boldsymbol{x}(t_0) + \int_{t_0}^{t_f} \left\{ \mathcal{H}(t) + \dot{\boldsymbol{\lambda}}^T(t)\boldsymbol{x}(t) \right\} dt$$

following integration of the $\boldsymbol{\lambda}^T \dot{\boldsymbol{x}}$ by parts, where

$$\Phi = \phi + \boldsymbol{\nu}^T \psi \tag{S11}$$

A variation of $\bar{J}$ writes

$$\delta \bar{J} = \left[(\Phi_x - \boldsymbol{\lambda}^T)\delta x\right]_{t=t_f} + \left[\boldsymbol{\lambda}^T \delta x\right]_{t=t_0} + \int_{t_0}^{t_f} \left[ \left(\mathcal{H}_x + \dot{\boldsymbol{\lambda}}^T\right) \delta x + \mathcal{H}_u \delta u \right] dt$$

for variations $\delta x(t)$ and $\delta u(t)$. The Lagrange mutlipliers are chosen so that the coefficients of $\delta x(t)$ and $\delta x(t_f)$ vanish

$$\dot{\boldsymbol{\lambda}}^T = -\mathcal{H}_x = -\boldsymbol{\lambda}^T f_x, \tag{S12}$$

with boundary conditions

$$\boldsymbol{\lambda}^T(t_f) = \phi_x(t_f) + \boldsymbol{\nu}^T \psi_x(t_f). \tag{S13}$$

For a stationary solution, $\delta \bar{J} = 0$ for arbitrary $\delta u(t)$, which implies

$$\mathcal{H}_u = \boldsymbol{\lambda}^T f_u = 0 \qquad t_0 \leq t \leq t_f. \tag{S14}$$

The problem defined by Eq. S1, Eq. S12, Eq. S13 and Eq. S14 is a two-point boundary value problem which can be solved by classical integration methods (Bryson 1999).

**Linear case**

In the linear case, the problem is a first-order linear dynamical system which can be solved explicitly. The solution consists in a $2n \times 2n$ matrix $\mathbf{D}(t)$ such that

$$\begin{pmatrix} \boldsymbol{x}(t) \\ \boldsymbol{\lambda}(t) \end{pmatrix} = \mathbf{D}(t)\mathcal{C} \tag{S15}$$

is the solution at time $t$, where $\mathcal{C} \in \mathbb{R}^{2n}$ is a vector determined by the boundary conditions (Eq. S3). To simplify we use $\psi\left[\bm{x}(t_f)\right] = \bm{x}(t_f) - \bm{x}_f$, but more complex boundary conditions can be handled as well (see below). To obtain $\mathcal{C}$, we write

$$\begin{pmatrix} \bm{x}_0 \\ \bm{\lambda}(t_0) \end{pmatrix} = \mathbf{D}(t_0)\mathcal{C} = \begin{pmatrix} \mathbf{D}_{11}(t_0) & \mathbf{D}_{12}(t_0) \\ \mathbf{D}_{21}(t_0) & \mathbf{D}_{22}(t_0) \end{pmatrix} \begin{pmatrix} \mathcal{C}_1 \\ \mathcal{C}_2 \end{pmatrix}$$

and

$$\begin{pmatrix} \bm{x}_f \\ \bm{\lambda}(t_f) \end{pmatrix} = \mathbf{D}(t_f)\mathcal{C} = \begin{pmatrix} \mathbf{D}_{11}(t_f) & \mathbf{D}_{12}(t_f) \\ \mathbf{D}_{21}(t_f) & \mathbf{D}_{22}(t_f) \end{pmatrix} \begin{pmatrix} \mathcal{C}_1 \\ \mathcal{C}_2 \end{pmatrix}.$$

Thus

$$\begin{pmatrix} \mathbf{D}_{11}(t_0) & \mathbf{D}_{12}(t_0) \\ \mathbf{D}_{11}(t_f) & \mathbf{D}_{12}(t_f) \end{pmatrix} \begin{pmatrix} \mathcal{C}_1 \\ \mathcal{C}_2 \end{pmatrix} = \begin{pmatrix} \bm{x}_0 \\ \bm{x}_f \end{pmatrix},$$

which gives

$$\mathcal{C} = \begin{pmatrix} \mathbf{D}_{11}(t_0) & \mathbf{D}_{12}(t_0) \\ \mathbf{D}_{11}(t_f) & \mathbf{D}_{12}(t_f) \end{pmatrix}^{-1} \begin{pmatrix} \bm{x}_0 \\ \bm{x}_f \end{pmatrix}. \tag{S16}$$

## Complete treatement of a linear case

Here we consider the problem of controlling an inertial point actuated by a linear muscle with a quadratic cost function. In Mayer formulation, the problem can be written

$$\begin{cases} \dot{x}_1 = u^2/2 \\ \dot{x}_2 = x_3 \\ \dot{x}_3 = x_4/m \\ \dot{x}_4 = (-x_4 + x_5)/\tau \\ \dot{x}_5 = (-x_5 + u)/\tau \end{cases}$$

The Hamiltonian (Eq. S10) writes

$$\begin{aligned} \mathcal{H} &= \lambda_1 u^2/2 + \lambda_2 x_3 + \lambda_3 x_4/m + \\ &\quad \lambda_4(-x_4 + x_5)/\tau + \lambda_5(-x_5 + u)/\tau \end{aligned}$$

from which the adjoint system (Eq. S12) can be obtained

$$\begin{aligned}
\dot{\lambda}_1 &= -\frac{\partial \mathcal{H}}{\partial x_1} = 0 \\
\dot{\lambda}_2 &= -\frac{\partial \mathcal{H}}{\partial x_2} = 0 \\
\dot{\lambda}_3 &= -\frac{\partial \mathcal{H}}{\partial x_3} = -\lambda_2 \\
\dot{\lambda}_4 &= -\frac{\partial \mathcal{H}}{\partial x_4} = -\lambda_3/m + \lambda_4/\tau \\
\dot{\lambda}_5 &= -\frac{\partial \mathcal{H}}{\partial x_5} = -\lambda_4/\tau + \lambda_5/\tau
\end{aligned}$$

The transversal condition (Eq. S14) is

$$\mathcal{H}_u = \lambda_1 u + \lambda_5/\tau = 0,$$

where $\lambda_1$ is a constant set at 1. The corresponding boundary value problem is

$$\begin{cases}
\dot{x}_2 = x_3 \\
\dot{x}_3 = x_4/m \\
\dot{x}_4 = (-x_4 + x_5)/\tau \\
\dot{x}_5 = (-x_5 - \lambda_5/\tau)/\tau \\
\dot{\lambda}_2 = 0 \\
\dot{\lambda}_3 = -\lambda_2 \\
\dot{\lambda}_4 = -\lambda_3/m + \lambda_4/\tau \\
\dot{\lambda}_5 = -\lambda_4/\tau + \lambda_5/\tau
\end{cases} \qquad (S17)$$

The constraints are defined by function $\Phi$ (Eq. S11) which can take different forms:

- Full constraints: position, velocity, activation, excitation

$$\begin{aligned}
\Phi &= x_1(t_f) + \nu_2 \left[x_2(t_f) - x_2^f\right] + \nu_3 \left[x_3(t_f) - x_3^f\right] + \\
&\quad \nu_4 \left[x_4(t_f) - x_4^f\right] + \nu_5 \left[x_5(t_f) - x_5^f\right]
\end{aligned}$$

- Partial constraints: position, velocity, activation

$$\begin{aligned}
\Phi &= x_1(t_f) + \nu_2 \left[x_2(t_f) - x_2^f\right] + \nu_3 \left[x_3(t_f) - x_3^f\right] + \\
&\quad \nu_4 \left[x_4(t_f) - x_4^f\right]
\end{aligned}$$

- Partial constraints: position, velocity

$$\Phi = x_1(t_f) + \nu_2 \left[x_2(t_f) - x_2^f\right] + \nu_3 \left[x_3(t_f) - x_3^f\right]$$

- Partial constraints: position

$$\Phi = x_1(t_f) + \nu_2 \left[x_2(t_f) - x_2^f\right]$$

The initial boundary conditions are

$$x_2(t_0) = x_2^0 \quad x_3(t_0) = x_3^0 \quad x_4(t_0) = x_4^0 \quad x_5(t_0) = x_5^0 \quad \text{(S18)}$$

The final boundary conditions are obtained using Eq. S13:

- Full constraints: position, velocity, activation, excitation

$$x_2(t_f) = x_2^f \quad x_3(t_f) = x_3^f \quad x_4(t_f) = x_4^f \quad x_5(t_f) = x_5^f \quad \text{(S19)}$$

$$\lambda_1(t_f) = 1 \quad \lambda_2(t_f) = \nu_2 \quad \lambda_3(t_f) = \nu_3 \quad \lambda_4(t_f) = \nu_4 \quad \lambda_5(t_f) = \nu_5 \quad \text{(S20)}$$

- Partial constraints: position, velocity, activation

$$x_2(t_f) = x_2^f \quad x_3(t_f) = x_3^f \quad x_4(t_f) = x_4^f \quad \text{(S21)}$$

$$\lambda_1(t_f) = 1 \quad \lambda_2(t_f) = \nu_2 \quad \lambda_3(t_f) = \nu_3 \quad \lambda_4(t_f) = \nu_4 \quad \lambda_5(t_f) = 0 \quad \text{(S22)}$$

- Partial constraints: position, velocity

$$x_2(t_f) = x_2^f \quad x_3(t_f) = x_3^f \quad \text{(S23)}$$

$$\lambda_1(t_f) = 1 \quad \lambda_2(t_f) = \nu_2 \quad \lambda_3(t_f) = \nu_3 \quad \lambda_4(t_f) = 0 \quad \lambda_5(t_f) = 0 \quad \text{(S24)}$$

- Partial constraints: position

$$x_2(t_f) = x_2^f \quad \text{(S25)}$$

$$\lambda_1(t_f) = 1 \quad \lambda_2(t_f) = \nu_2 \quad \lambda_3(t_f) = 0 \quad \lambda_4(t_f) = 0 \quad \lambda_5(t_f) = 0 \quad \text{(S26)}$$

From Eq. S27, the solution consists in a $2n \times 2n$ matrix $\mathbf{D}(t)$ ($n = 4$) such that

$$\begin{pmatrix} \boldsymbol{x}(t) \\ \boldsymbol{\lambda}(t) \end{pmatrix} = \mathbf{D}(t)\mathcal{C} \tag{S27}$$

is the solution at time $t$, where $\mathcal{C} \in \mathbb{R}^{2n}$ is a vector determined by the initial and final boundary conditions. Here $\mathbf{D}$ is the solution to the boundary value problem (Eq. S17), which can be obtained explicitly using tools of symbolic calculus.

To obtain $\mathcal{C}$, we write

$$\begin{pmatrix} \boldsymbol{x}_0 \\ \boldsymbol{\lambda}(t_0) \end{pmatrix} = \mathbf{D}(t_0)\mathcal{C} = \mathbf{D}^0 \mathcal{C}$$

and

$$\begin{pmatrix} \boldsymbol{x}_f \\ \boldsymbol{\lambda}(t_f) \end{pmatrix} = \mathbf{D}(t_f)\mathcal{C} = \mathbf{D}^f \mathcal{C}$$

and we extract what is known from these relationships in the different cases (full constraints: Eq. S19 and Eq. S20; partial constraints on position, velocity, activation: Eq. S21 and Eq. S22; partial constraints on position, velocity: Eq. S23 and Eq. S24; partial constraints on position: Eq. S25 and Eq. S26).

We obtain a relationship

$$\mathbf{M}\boldsymbol{q} = \boldsymbol{p} \tag{S28}$$

where $\mathbf{M}$ contains elements of $\mathbf{D}^0$ and $\mathbf{D}^f$, $\boldsymbol{q}$ the vector $\mathcal{C}$ and some elements of $\boldsymbol{\nu}$, and $\boldsymbol{p}$ the vector $\boldsymbol{x}_0$ and some elements of $\boldsymbol{x}_f$. Taking $\boldsymbol{q} = \mathbf{M}^{-1}\boldsymbol{p}$ gives the vector $\mathcal{C}$.

For the case of full constraints (position, velocity, activation, excitation), there are 8 unknowns (8 in $\mathcal{C}$). We get 4 equations for $\boldsymbol{x}_0$ (Eq. S18), and 4 equations for $\boldsymbol{x}_f$ (Eq. S19), and Eq. S28 becomes

$$\begin{pmatrix} \mathbf{D}^0_{11} & \mathbf{D}^0_{12} & \mathbf{D}^0_{13} & \mathbf{D}^0_{14} & \mathbf{D}^0_{15} & \mathbf{D}^0_{16} & \mathbf{D}^0_{17} & \mathbf{D}^0_{18} \\ \mathbf{D}^0_{21} & \mathbf{D}^0_{22} & \mathbf{D}^0_{23} & \mathbf{D}^0_{24} & \mathbf{D}^0_{25} & \mathbf{D}^0_{26} & \mathbf{D}^0_{27} & \mathbf{D}^0_{28} \\ \mathbf{D}^0_{31} & \mathbf{D}^0_{32} & \mathbf{D}^0_{33} & \mathbf{D}^0_{34} & \mathbf{D}^0_{35} & \mathbf{D}^0_{36} & \mathbf{D}^0_{37} & \mathbf{D}^0_{38} \\ \mathbf{D}^0_{41} & \mathbf{D}^0_{42} & \mathbf{D}^0_{43} & \mathbf{D}^0_{44} & \mathbf{D}^0_{45} & \mathbf{D}^0_{46} & \mathbf{D}^0_{47} & \mathbf{D}^0_{48} \\ \mathbf{D}^f_{11} & \mathbf{D}^f_{12} & \mathbf{D}^f_{13} & \mathbf{D}^f_{14} & \mathbf{D}^f_{15} & \mathbf{D}^f_{16} & \mathbf{D}^f_{17} & \mathbf{D}^f_{18} \\ \mathbf{D}^f_{21} & \mathbf{D}^f_{22} & \mathbf{D}^f_{23} & \mathbf{D}^f_{24} & \mathbf{D}^f_{25} & \mathbf{D}^f_{26} & \mathbf{D}^f_{27} & \mathbf{D}^f_{28} \\ \mathbf{D}^f_{31} & \mathbf{D}^f_{32} & \mathbf{D}^f_{33} & \mathbf{D}^f_{34} & \mathbf{D}^f_{35} & \mathbf{D}^f_{36} & \mathbf{D}^f_{37} & \mathbf{D}^f_{38} \\ \mathbf{D}^f_{41} & \mathbf{D}^f_{42} & \mathbf{D}^f_{43} & \mathbf{D}^f_{44} & \mathbf{D}^f_{45} & \mathbf{D}^f_{46} & \mathbf{D}^f_{47} & \mathbf{D}^f_{48} \end{pmatrix} \begin{pmatrix} \mathcal{C}_1 \\ \mathcal{C}_2 \\ \mathcal{C}_3 \\ \mathcal{C}_4 \\ \mathcal{C}_5 \\ \mathcal{C}_6 \\ \mathcal{C}_7 \\ \mathcal{C}_8 \end{pmatrix} = \begin{pmatrix} x_2^0 \\ x_3^0 \\ x_4^0 \\ x_5^0 \\ x_2^f \\ x_3^f \\ x_4^f \\ x_5^f \end{pmatrix}$$

For the case of partial constraints on position, velocity, and activation, there are 11 unknowns (8 in $\mathcal{C}$, $\nu_2$, $\nu_3$, $\nu_4$). We get 4 equations for $\boldsymbol{x}_0$ (Eq. S18), 3 equations for $\boldsymbol{x}_f$ (Eq. S21), 4 equations for $\boldsymbol{\lambda}(t_f)$ (Eq. S22), and Eq. S28 becomes

$$\begin{pmatrix} \mathbf{D}^0_{11} & \mathbf{D}^0_{12} & \mathbf{D}^0_{13} & \mathbf{D}^0_{14} & \mathbf{D}^0_{15} & \mathbf{D}^0_{16} & \mathbf{D}^0_{17} & \mathbf{D}^0_{18} & 0 & 0 & 0 \\ \mathbf{D}^0_{21} & \mathbf{D}^0_{22} & \mathbf{D}^0_{23} & \mathbf{D}^0_{24} & \mathbf{D}^0_{25} & \mathbf{D}^0_{26} & \mathbf{D}^0_{27} & \mathbf{D}^0_{28} & 0 & 0 & 0 \\ \mathbf{D}^0_{31} & \mathbf{D}^0_{32} & \mathbf{D}^0_{33} & \mathbf{D}^0_{34} & \mathbf{D}^0_{35} & \mathbf{D}^0_{36} & \mathbf{D}^0_{37} & \mathbf{D}^0_{38} & 0 & 0 & 0 \\ \mathbf{D}^0_{41} & \mathbf{D}^0_{42} & \mathbf{D}^0_{43} & \mathbf{D}^0_{44} & \mathbf{D}^0_{45} & \mathbf{D}^0_{46} & \mathbf{D}^0_{47} & \mathbf{D}^0_{48} & 0 & 0 & 0 \\ \mathbf{D}^f_{11} & \mathbf{D}^f_{12} & \mathbf{D}^f_{13} & \mathbf{D}^f_{14} & \mathbf{D}^f_{15} & \mathbf{D}^f_{16} & \mathbf{D}^f_{17} & \mathbf{D}^f_{18} & 0 & 0 & 0 \\ \mathbf{D}^f_{21} & \mathbf{D}^f_{22} & \mathbf{D}^f_{23} & \mathbf{D}^f_{24} & \mathbf{D}^f_{25} & \mathbf{D}^f_{26} & \mathbf{D}^f_{27} & \mathbf{D}^f_{18} & 0 & 0 & 0 \\ \mathbf{D}^f_{31} & \mathbf{D}^f_{32} & \mathbf{D}^f_{33} & \mathbf{D}^f_{34} & \mathbf{D}^f_{35} & \mathbf{D}^f_{36} & \mathbf{D}^f_{37} & \mathbf{D}^f_{38} & 0 & 0 & 0 \\ \mathbf{D}^f_{51} & \mathbf{D}^f_{52} & \mathbf{D}^f_{53} & \mathbf{D}^f_{54} & \mathbf{D}^f_{55} & \mathbf{D}^f_{56} & \mathbf{D}^f_{57} & \mathbf{D}^f_{58} & -1 & 0 & 0 \\ \mathbf{D}^f_{61} & \mathbf{D}^f_{62} & \mathbf{D}^f_{63} & \mathbf{D}^f_{64} & \mathbf{D}^f_{65} & \mathbf{D}^f_{66} & \mathbf{D}^f_{67} & \mathbf{D}^f_{68} & 0 & -1 & 0 \\ \mathbf{D}^f_{71} & \mathbf{D}^f_{72} & \mathbf{D}^f_{73} & \mathbf{D}^f_{74} & \mathbf{D}^f_{75} & \mathbf{D}^f_{76} & \mathbf{D}^f_{77} & \mathbf{D}^f_{78} & 0 & 0 & -1 \\ \mathbf{D}^f_{81} & \mathbf{D}^f_{82} & \mathbf{D}^f_{83} & \mathbf{D}^f_{84} & \mathbf{D}^f_{85} & \mathbf{D}^f_{86} & \mathbf{D}^f_{87} & \mathbf{D}^f_{88} & 0 & 0 & 0 \end{pmatrix} \begin{pmatrix} \mathcal{C}_1 \\ \mathcal{C}_2 \\ \mathcal{C}_3 \\ \mathcal{C}_4 \\ \mathcal{C}_5 \\ \mathcal{C}_6 \\ \mathcal{C}_7 \\ \mathcal{C}_8 \\ \nu_2 \\ \nu_3 \\ \nu_4 \end{pmatrix} = \begin{pmatrix} x^0_2 \\ x^0_3 \\ x^0_4 \\ x^0_5 \\ x^f_2 \\ x^f_3 \\ x^f_4 \\ 0 \\ 0 \\ 0 \\ 0 \end{pmatrix}$$

For the case of partial constraints on position and velocity, there are 10 unknowns (8 in $\mathcal{C}$, $\nu_2$, $\nu_3$). We get 4 equations for $x_0$ (Eq. S18), 2 equations for $x_f$ (Eq. S23), 4 equations for $\lambda(t_f)$ (Eq. S24), and Eq. S28 becomes

$$\begin{pmatrix} \mathbf{D}^0_{11} & \mathbf{D}^0_{12} & \mathbf{D}^0_{13} & \mathbf{D}^0_{14} & \mathbf{D}^0_{15} & \mathbf{D}^0_{16} & \mathbf{D}^0_{17} & \mathbf{D}^0_{18} & 0 & 0 \\ \mathbf{D}^0_{21} & \mathbf{D}^0_{22} & \mathbf{D}^0_{23} & \mathbf{D}^0_{24} & \mathbf{D}^0_{25} & \mathbf{D}^0_{26} & \mathbf{D}^0_{27} & \mathbf{D}^0_{28} & 0 & 0 \\ \mathbf{D}^0_{31} & \mathbf{D}^0_{32} & \mathbf{D}^0_{33} & \mathbf{D}^0_{34} & \mathbf{D}^0_{35} & \mathbf{D}^0_{36} & \mathbf{D}^0_{37} & \mathbf{D}^0_{38} & 0 & 0 \\ \mathbf{D}^0_{41} & \mathbf{D}^0_{42} & \mathbf{D}^0_{43} & \mathbf{D}^0_{44} & \mathbf{D}^0_{45} & \mathbf{D}^0_{46} & \mathbf{D}^0_{47} & \mathbf{D}^0_{48} & 0 & 0 \\ \mathbf{D}^f_{11} & \mathbf{D}^f_{12} & \mathbf{D}^f_{13} & \mathbf{D}^f_{14} & \mathbf{D}^f_{15} & \mathbf{D}^f_{16} & \mathbf{D}^f_{17} & \mathbf{D}^f_{18} & 0 & 0 \\ \mathbf{D}^f_{21} & \mathbf{D}^f_{22} & \mathbf{D}^f_{23} & \mathbf{D}^f_{24} & \mathbf{D}^f_{25} & \mathbf{D}^f_{26} & \mathbf{D}^f_{27} & \mathbf{D}^f_{18} & 0 & 0 \\ \mathbf{D}^f_{51} & \mathbf{D}^f_{52} & \mathbf{D}^f_{53} & \mathbf{D}^f_{54} & \mathbf{D}^f_{55} & \mathbf{D}^f_{56} & \mathbf{D}^f_{57} & \mathbf{D}^f_{58} & -1 & 0 \\ \mathbf{D}^f_{61} & \mathbf{D}^f_{62} & \mathbf{D}^f_{63} & \mathbf{D}^f_{64} & \mathbf{D}^f_{65} & \mathbf{D}^f_{66} & \mathbf{D}^f_{67} & \mathbf{D}^f_{68} & 0 & -1 \\ \mathbf{D}^f_{71} & \mathbf{D}^f_{72} & \mathbf{D}^f_{73} & \mathbf{D}^f_{74} & \mathbf{D}^f_{75} & \mathbf{D}^f_{76} & \mathbf{D}^f_{77} & \mathbf{D}^f_{78} & 0 & 0 \\ \mathbf{D}^f_{81} & \mathbf{D}^f_{82} & \mathbf{D}^f_{83} & \mathbf{D}^f_{84} & \mathbf{D}^f_{85} & \mathbf{D}^f_{86} & \mathbf{D}^f_{87} & \mathbf{D}^f_{88} & 0 & 0 \end{pmatrix} \begin{pmatrix} \mathcal{C}_1 \\ \mathcal{C}_2 \\ \mathcal{C}_3 \\ \mathcal{C}_4 \\ \mathcal{C}_5 \\ \mathcal{C}_6 \\ \mathcal{C}_7 \\ \mathcal{C}_8 \\ \nu_2 \\ \nu_3 \end{pmatrix} = \begin{pmatrix} x^0_2 \\ x^0_3 \\ x^0_4 \\ x^0_5 \\ x^f_2 \\ x^f_3 \\ 0 \\ 0 \\ 0 \\ 0 \end{pmatrix}$$

For the case of partial constraints on position, there are 9 unknowns (8 in $\mathcal{C}$, $\nu_2$). We get 4 equations for $x_0$ (Eq. S18), 1 equation for $x_f$ (Eq. S25), 4 equations for $\lambda(t_f)$ (Eq. S26), and Eq. S28 becomes

$$\begin{pmatrix} \mathbf{D}^0_{11} & \mathbf{D}^0_{12} & \mathbf{D}^0_{13} & \mathbf{D}^0_{14} & \mathbf{D}^0_{15} & \mathbf{D}^0_{16} & \mathbf{D}^0_{17} & \mathbf{D}^0_{18} & 0 \\ \mathbf{D}^0_{21} & \mathbf{D}^0_{22} & \mathbf{D}^0_{23} & \mathbf{D}^0_{24} & \mathbf{D}^0_{25} & \mathbf{D}^0_{26} & \mathbf{D}^0_{27} & \mathbf{D}^0_{28} & 0 \\ \mathbf{D}^0_{31} & \mathbf{D}^0_{32} & \mathbf{D}^0_{33} & \mathbf{D}^0_{34} & \mathbf{D}^0_{35} & \mathbf{D}^0_{36} & \mathbf{D}^0_{37} & \mathbf{D}^0_{38} & 0 \\ \mathbf{D}^0_{41} & \mathbf{D}^0_{42} & \mathbf{D}^0_{43} & \mathbf{D}^0_{44} & \mathbf{D}^0_{45} & \mathbf{D}^0_{46} & \mathbf{D}^0_{47} & \mathbf{D}^0_{48} & 0 \\ \mathbf{D}^f_{11} & \mathbf{D}^f_{12} & \mathbf{D}^f_{13} & \mathbf{D}^f_{14} & \mathbf{D}^f_{15} & \mathbf{D}^f_{16} & \mathbf{D}^f_{17} & \mathbf{D}^f_{18} & 0 \\ \mathbf{D}^f_{51} & \mathbf{D}^f_{52} & \mathbf{D}^f_{53} & \mathbf{D}^f_{54} & \mathbf{D}^f_{55} & \mathbf{D}^f_{56} & \mathbf{D}^f_{57} & \mathbf{D}^f_{58} & -1 \\ \mathbf{D}^f_{61} & \mathbf{D}^f_{62} & \mathbf{D}^f_{63} & \mathbf{D}^f_{64} & \mathbf{D}^f_{65} & \mathbf{D}^f_{66} & \mathbf{D}^f_{67} & \mathbf{D}^f_{68} & 0 \\ \mathbf{D}^f_{71} & \mathbf{D}^f_{72} & \mathbf{D}^f_{73} & \mathbf{D}^f_{74} & \mathbf{D}^f_{75} & \mathbf{D}^f_{76} & \mathbf{D}^f_{77} & \mathbf{D}^f_{78} & 0 \\ \mathbf{D}^f_{81} & \mathbf{D}^f_{82} & \mathbf{D}^f_{83} & \mathbf{D}^f_{84} & \mathbf{D}^f_{85} & \mathbf{D}^f_{86} & \mathbf{D}^f_{87} & \mathbf{D}^f_{88} & 0 \end{pmatrix} \begin{pmatrix} \mathcal{C}_1 \\ \mathcal{C}_2 \\ \mathcal{C}_3 \\ \mathcal{C}_4 \\ \mathcal{C}_5 \\ \mathcal{C}_6 \\ \mathcal{C}_7 \\ \mathcal{C}_8 \\ \nu_2 \end{pmatrix} = \begin{pmatrix} x^0_2 \\ x^0_3 \\ x^0_4 \\ x^0_5 \\ x^f_2 \\ 0 \\ 0 \\ 0 \\ 0 \end{pmatrix}$$